\documentclass[aps,pre,twocolumn,superscriptaddress,amsmath,showpacs,amssymb,nofootinbib]{revtex4-2}
\usepackage{hyperref}
\usepackage{graphicx}
\usepackage{enumerate}
\usepackage{dcolumn}
\usepackage{bm}

\usepackage{subfigure}
\usepackage{siunitx}
\usepackage{todonotes}
\usepackage[mathscr]{eucal}

\presetkeys{todonotes}{inline}{}

\usepackage{booktabs}

\usepackage[sort&compress]{natbib} 		   

\newcommand{\df}{\mathrm{d}}
\newcommand{\vv}[1]{\bm{#1}}
\newcommand{\te}[1]{\mathrm{#1}}

\newcommand{\ie}{i.e.}
\newcommand{\eg}{e.g.}

\newcommand{\kB}{{k_\te B}}
\newcommand{\Dr}{{D_\te r}}
\newcommand{\Deff}{{D_\te{eff}}}
\newcommand{\vA}{{v_\te A}}
\newcommand{\va}{{v_\te a}}
\newcommand{\vn}{{v_\te n}}
\newcommand{\Pa}{{p_\te a}}
\newcommand{\Pp}{{p_\te p}}
\newcommand{\RA}{{\rho_\te A}}
\newcommand{\Ra}{{\rho_\te a}}
\newcommand{\Rp}{{\rho_\te p}}
\newcommand{\lA}{{\lambda_{\te A}}}
\newcommand{\la}{{\lambda_{\te a}}}
\newcommand{\lp}{{\lambda_{\te p}}}
\newcommand{\lone}{{\lambda_{\te{n}_1}^{-1}}}
\newcommand{\ltwo}{{\lambda_{\te{n}_2}^{-1}}}
\newcommand{\lthree}{{\lambda_{\te{n}_3}^{-1}}}
\newcommand{\Ca}{{C_\te a}}
\newcommand{\Cp}{{C_\te p}}
\newcommand{\xif}{{x_\te{if}}}

\newcommand{\hn}{{\vv{\hat n}}}

\newcommand{\mynabla}{{\boldsymbol{\nabla}}}

\newcommand\uleipE{\affiliation{
Peter Debye Institute for Soft Matter Physics, 
Leipzig University, 04103 Leipzig, Germany}}
\newcommand\uleipT{\affiliation{
Institute for Theoretical Physics, 
Leipzig University, 
04103 Leipzig, Germany}}
\newcommand\chau{\affiliation{ 
 Charles University,  
 Faculty of Mathematics and Physics, 
 V Hole{\v s}ovi{\v c}k{\' a}ch 2, 
 CZ-180~00~Praha, Czech Republic 
}}

\DeclareSIUnit\length{\text{$\sqrt{D/2k}$}}
\DeclareSIUnit\lengthcont{\text{$\sqrt{D/\Dr}$}}

\begin{document}

\preprint{APS/123-QED}

\title{Polarization-Density Patterns of Active Particles in Motility Gradients}

\author{Sven Auschra}
\email{sven.auschra@itp.uni-leipzig.de}
\uleipT
\author{Viktor Holubec}
\email{viktor.holubec@mff.cuni.cz}
\uleipT
\chau
\author{Nicola Andreas Söker}
\email{ns66qoxa@.uni-leipzig.de}
\uleipE
\author{Frank Cichos}\email{cichos@uni-leipzig.de}\uleipE
\author{Klaus Kroy}
\email{klaus.kroy@itp.uni-leipzig.de}\uleipT

\date{\today}

\begin{abstract}
  The co-localization of density modulations and particle polarization is a characteristic emergent feature of motile active matter in activity gradients. It can therefore play the role of a smoking gun for the mesoscale detection of intrinsic microscopic activity.
  We employ the active-Brownian-particle (ABP) model to derive precise analytical expressions for the density and polarization profiles of a single Janus-type swimmer in the vicinity of an abrupt activity step, including situations with an orientation-dependent propulsion speed. Our results agree well with measurement data for a hot microswimmer presented in the companion paper \cite{Soeker2020ActivityFieldsPRL} and can serve as a template for more complex applications, e.g., to motility-induced phase separation or studies of physical boundaries.
 The essential physics behind our formal results is robustly captured and elucidated by a  schematic two-species ``run-and-tumble'' model.
\end{abstract}

\maketitle

\section{Introduction}
\label{sec:introduction}
The surging field  of active matter \cite{Ramaswamy2010TheMatter,Cates2012DiffusivePhysics,romanczuk2012} aims for a microscopic understanding and control of the material properties of assemblies of interacting energy-consuming elements  \cite{Bechinger2016ActiveEnvironments}. In particular, examples for \emph{motile} active matter are ubiquitous in nature, ranging from flocks of birds \cite{vicsek2012CollMotion}, via swarms of insects \cite{sponberg2017animalLocomotion} to colonies of bacteria, such as Escherichia coli \cite{berg04Ecoli}. The wealth of observed natural phenomena  has stimulated many laboratory studies of artificial active fluids of suspended inanimate microswimmers (\cite{Bechinger2016ActiveEnvironments}, Tab.~I). Such ``active-particle systems''  often consist of simple colloidal particles propelled by a form of self-phoresis  \cite{Anderson1989ColloidForces,Jiang2010ActiveBeam,Falasco2016ExactSwimmer,Buttinoni2012ActiveLight,yang2013thermophorFlowFields,moran2010electrophoresis,golestanian2005Diffusophoresis,Howse2007Self-MotileWalk}. Numerous interesting features have been observed already on the level of a single or a few active particles \cite{Falasco2016ExactSwimmer,shen2018WallInteract,bayati2019DynamicsNearWall,Bickel2014PolarizationParticles,geiseler2017ActivityWaves,olarte_plata2020PolarJPExtField,saha2019PairingWaltzingScattering,babak2020InteractingJPgeneric}, which open  a wide range of potential applications \cite{Qian2013HarnessingNudging,Selmke2018TheoryTransport,Selmke2018TheoryConfinement,Volpe2011MicroswimmersEnvironments,takagi2014obstacles,Das2015BoundariesSpheres,uspal2016guidepatterns,palagi2016lightstruct,Simmchen2016TopographicalMicroswimmers,baraban2012magnsteer,burdick2008magnsteer,ahmed2013magnsteer}. Microswimmers moreover exhibit rich collective dynamics, ranging from mesoscopic turbulence via collective oscillations to macroscopic motility-induced phase separation (MIPS) \cite{wensink2012mesoscaleTurb,grossmann2014mesoscaleTurb,saha2014clusterAtersOscill,bialke2013PhaseSep,Cates2013WhenSeparation,speck2014Cahn-Hilliard,wysocki2014denseSusp,Zottl2014HydrodynamicsConfinement,Cates2015MIPS,grossmann2015patternForm,Mijalkov2016EngineeringBehaviors,takatori2016review,caprini2020velocityAlign}.

\begin{figure}[tb!]
  \centering
  \includegraphics[width=\columnwidth]{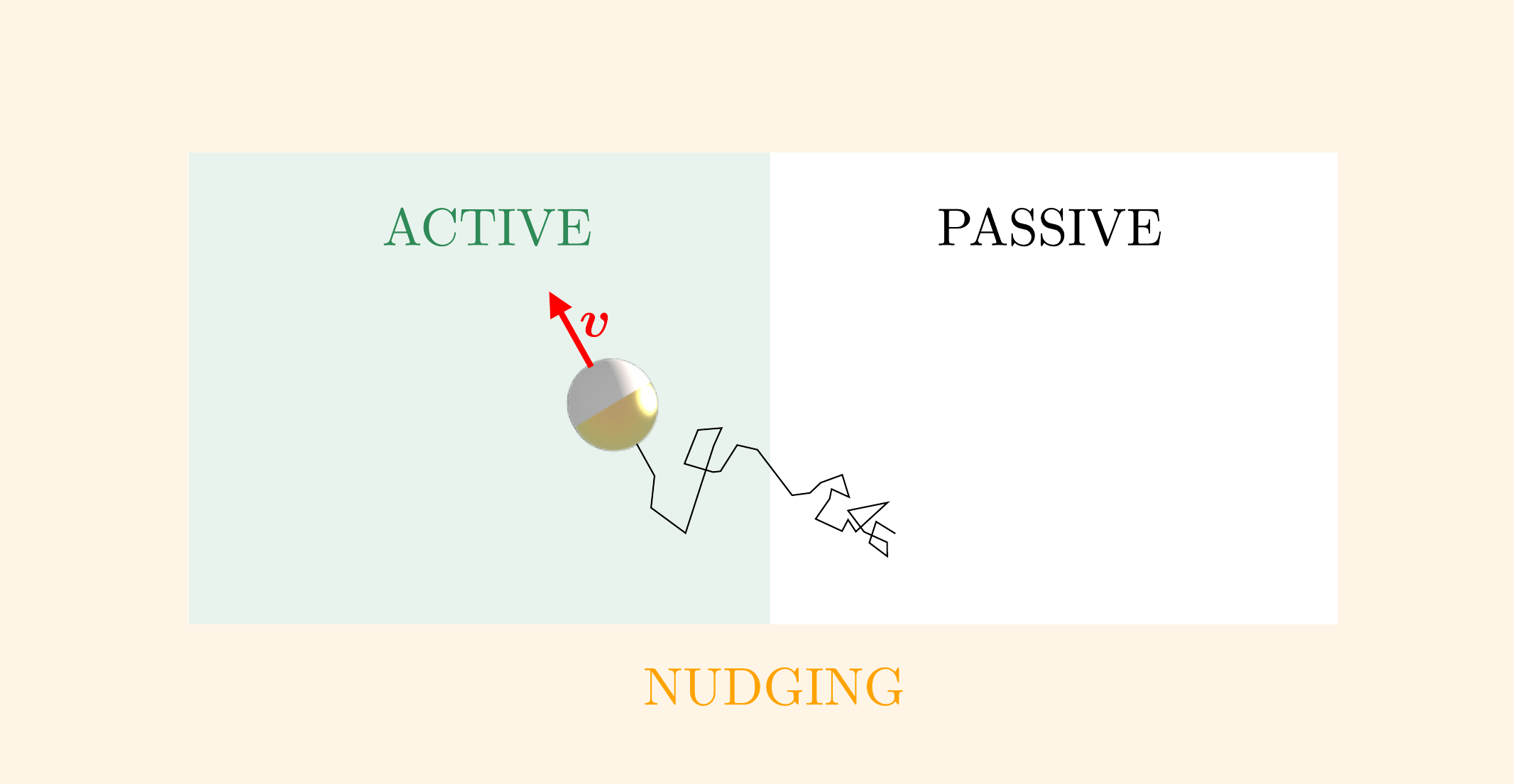}
  \caption{Sketch of the experimental setup studied in the companion article~\cite{Soeker2020ActivityFieldsPRL} (not to scale). Photon nudging \cite{Selmke2018TheoryTransport,Selmke2018TheoryConfinement} is employed to confine a self-thermophoretic Janus swimmer within a rectangular arena without physical boundaries or confining forces. Its in-plane orientation and position are observed by microscopy \cite{Soeker2020ActivityFieldsPRL}. Upon entering from the PASSIVE into the ACTIVE region, the particle's Browian motion gets boosted by self-propulsion  with velocity $\boldsymbol v$ along its symmetry axis.}
  \label{fig:setup_exp}
\end{figure}

A paradigmatic problem for a single active particle, namely its motion in a heterogeneous activity/motility field, was experimentally studied in the companion article \cite{Soeker2020ActivityFieldsPRL}. 
The crux of the experimental setup is that it allows for long-time observations of a single autonomous microswimmer \cite{Gompper2016MicroswimmersBehavior} near an abrupt activity step. The latter may be thought of as a concomitant feature of most physical boundaries (\eg, sedimentation   \cite{hermann2018Sediment,vachier2019SedimentABP,enculescu2011PolarOrderGrav,ginot2018SedimentatPolPress}, wall adsorption  \cite{hermann2020PolStateFct,Speck2016IdealBulkPressure,Wagner2017ABPsUnderConfinement,Elgeti2013WallAccumulation}), and even of collective phenomena such as MIPS interface formation  \cite{hermann2020PolStateFct,paliwal2018ChemPot,hermann2019MIPS,Solon2018GeneralizedMatter,solon2018GenTDofMIPS,prymidis2016VapLiquCoeyInLJSys,paliwal2017SurfTensionLJSys}. Importantly, the setup, schematically sketched in Fig.~\ref{fig:setup_exp},  confines the active particle to a planar arena by photon nudging \cite{Selmke2018TheoryTransport,Selmke2018TheoryConfinement}, hence without imposing any lateral physical boundaries or confinement forces. It thereby enables the experimental study of the emerging interfacial patterns at the central activity step without any of the fundamentally unrelated perturbations usually encountered in practical applications. 
In many ways, the setup can thus be likened to the idealized textbook quantum-mechanics problem of a particle in a potential well. The main finding in Ref.~\cite{Soeker2020ActivityFieldsPRL} is an emerging characteristic polarization-density pattern in an interfacial layer around the motility step, which can serve as a distinctive trait of active versus passive Brownian particle motion. And the main tasks of the present contribution are its precise theoretical computation and the discussion of its physical implications.  

To this end, we employ the (standard) active-Brownian-particle (ABP) model \cite{erdmann2000ABP,schweitzer2007brownianagents,romanczuk2012,Cates2013WhenSeparation,solon2015} for a single motile spherical particle.  Our analytical theory transcends and improves previous literature results \cite{Schnitzer1993TheoryChemotaxis,Malakar2018RunAndTumble1D,Sharma2017BrownianActivity,Fischer2020Quorum-sensingMotility,fischer2020erratum}, and suggests itself as a practical numerical tool for the approximate reconstruction of the full picture from incomplete and coarse-grained active-particle data for more complex geometries and interacting many-body problems.

\section{General Theory}
\label{sec:theory}

\subsection{Moment Equations}
\label{sec:general}

\begin{figure}[tb!]
  \centering
 \includegraphics[width=\columnwidth]{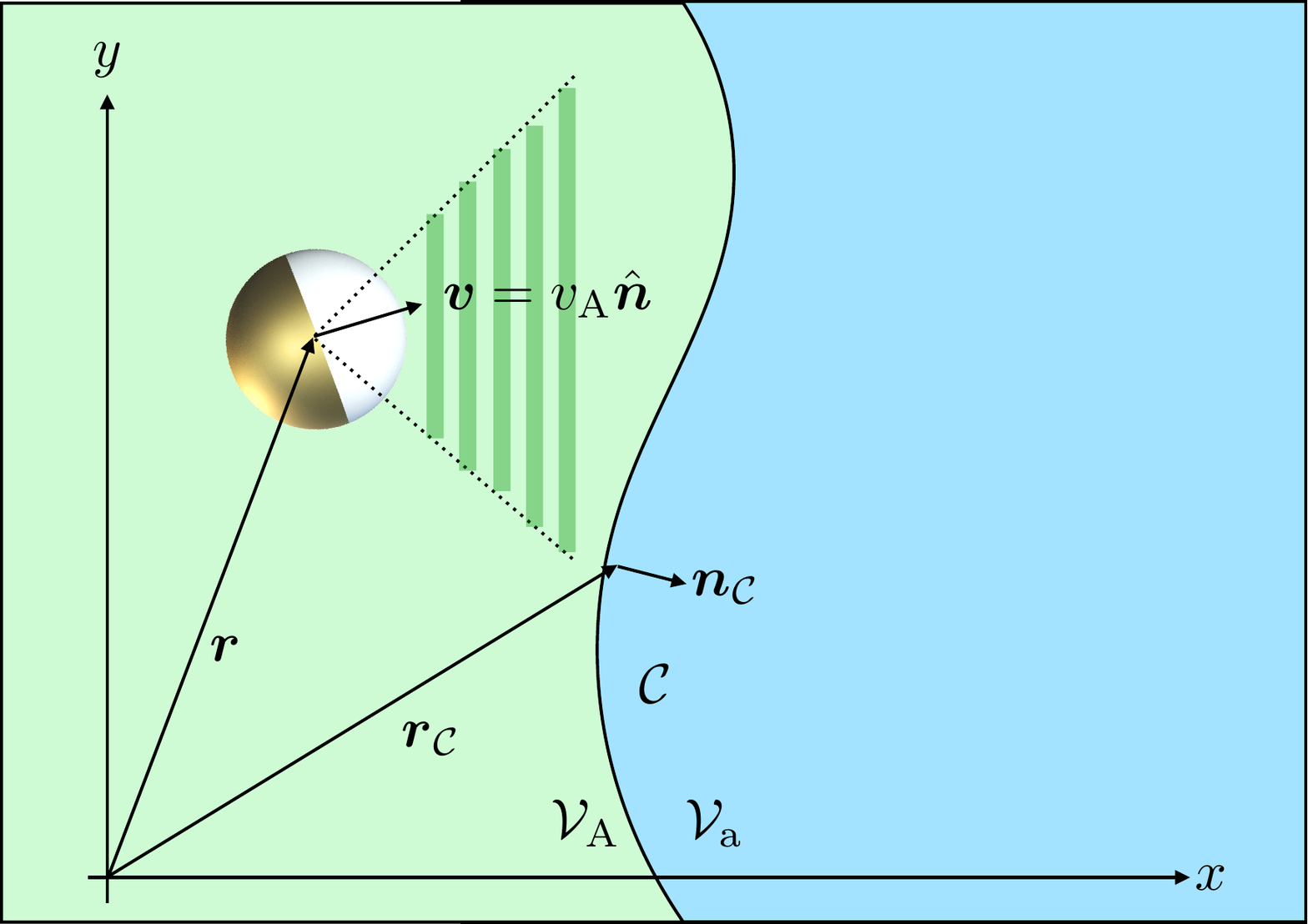}
 \caption{A Janus particle with orientation $\hn$ and position $\vv r$.
   The planar arena is divided into two sub-regions $\mathcal V_{\te{A},\te{a}}$ with distinct propulsion speeds, $v_{\te{A},\te{a}}$.  An optionally restricted acceptance range of particle orientations (shaded area) allows us to account for photon nudging.
   The interface $\mathcal C$ between the regions $\mathcal V_{\te{A},\te{a}}$ is parametrized by its position vector $\vv r_{\mathcal C}$ and  normal vector $\vv n_{\mathcal C}$ (pointing towards $\mathcal V_\te{a}$). }
  \label{fig:setup_2d_general}
\end{figure}

We idealize the experimental thermophoretic microswimmer by the standard ABP model \cite{erdmann2000ABP,schweitzer2007brownianagents,romanczuk2012,Cates2013WhenSeparation,solon2015} of an overdamped particle, whose propulsion speed $v(\vv r, \hn)$ depends on its position $\vv r$ and optionally also on its orientation $\hn$, according to the Langevin equations
\begin{equation}
  \label{eq:langevin_general}
  \partial_t \vv r
  =
  v(\vv r, \hn) \hn
  +
  \sqrt{2D} \boldsymbol{\xi}_{\te t},
  \qquad
  \partial_t \hn
  =
  \sqrt{2\Dr} \boldsymbol{\xi}_{\te r}
  \times
  \hn.
\end{equation}
Here, $D$ and $\Dr$ are the diffusion coefficients corresponding to the independent, unit variance, unbiased Gaussian white noise processes
$\boldsymbol{\xi}_{\te t,r}(t)$ pertaining to the particle's translation and rotation, respectively.
Some notation is illustrated by Fig.~\ref{fig:setup_2d_general} for a piece-wise constant activity profile in a planar setup.

The time evolution of the dynamic probability density $f(\vv r,\hn, t)$ for finding the Janus swimmer at time $t$ at position $\vv r$ with the in-plane orientation $\hn$ is described by the Fokker-Planck equation (FPE) \cite{golestanian2012CollectiveBehav,Cates2013WhenSeparation,solon2015}
\begin{equation}
  \label{eq:FPE_full}
  \partial_t f
  =
  D \mynabla^2 f
  +
  \Dr \mynabla_n^2 f
  -
  \mynabla \cdot [f v(\vv r, \hn) \hn],
\end{equation}
for dimensionality $d=2,3$.
Here, $\partial_t$ denotes the partial time derivative, and $\mynabla^2$ and $\mynabla_n^2$ are the translational and rotational parts of the Laplacian acting on $\vv r$ and $\hn$, respectively.
To extract measurable predictions from the  model, we truncate the (exact) moment expansion of $f$ with respect to the orientation $\hn$ \cite{bertin2006BoltzmannHydro, golestanian2012CollectiveBehav,Cates2013WhenSeparation}:
\begin{equation}
  \label{eq:moment_expansion_general}
    f(\vv r, \hn, t)
    =
    \frac{1}{\mathcal S_d}
    \left[
      \rho(\vv r, t)
      +
      d \vv p(\vv r, t)
      \cdot \hn
  \right].
\end{equation}
Here we used the following abbreviations
\begin{align}
  \mathcal S_d
  &\equiv
    \int \df \hn
    \quad \hspace{1.3cm}
    \text{(unit sphere surface)}
  \\    
  \label{eq:def_density}
  \rho(\vv r)
  &\equiv
    \int \df \hn ~
    f(\vv r, \hn)
    \quad \hspace{0.15cm}
    \text{(particle density)}
  \\
  \label{eq:def_polarization}
  \vv p(\vv r)
  &\equiv
    \int \df \hn ~
    \hn f(\vv r, \hn)
    \quad 
    \text{(polarization density)}.
\end{align}
Motivated by the experimental setup, the activity profile is modelled as $v(\vv r, \hn) = v(\vv r) \chi_{\mathcal A}(\hn)$.  Here, $v(\vv r)$ reflects the occurrence of inhomogeneous activity in position space, and the indicator function $\chi_{\mathcal A}(\hn)$, which is unity if $\hn \in \mathcal A \subseteq \mathcal S_d$ and zero otherwise, accounts for a limitation of the activity to an acceptance range $\mathcal A$ of the particle orientation. This enables us to include the effect of photon nudging.
We also note that the truncation in Eq.~(\ref{eq:moment_expansion_general}) is not systematic with respect to a small parameter but physically motivated. Namely, the moment equations for $\rho(x)$ and $p(x)$ derived from it in Sec.~\ref{sec:plan-active-pass} for one-dimensional activity profiles $v(x)$ can be mapped onto an exact description of a one-dimensional run-and-tumble model, which captures the relevant physics for arbitrary activity profiles and dimensions (see Sec.~\ref{sec:2-species-model}).

Multiplying Eq.~\eqref{eq:FPE_full} by 1 or $\hn$, using
\(
\mynabla_n^2 \vv n = -(d-1) \vv n 
\), and integrating the result over the orientational degrees of freedom, yields the two moment equations~\cite{Cates2012DiffusivePhysics}
\begin{align}
  \label{eq:mom_equ_rho_general}
  \partial_t \rho(\vv r, t)
  &=
    - \mynabla
    \cdot
    \vv J(\vv r, t),
  \\[0.5em]
  \label{eq:mom_equ_pol_general}
  \partial_t \vv p(\vv r, t)
  &=
    -(d-1)\Dr \vv p(\vv r, t)
    -
    \mynabla
    \cdot
    \vv M(\vv r, t).
\end{align}
Here, we introduced the (orientation-averaged) flux
\begin{equation}
  \label{eq:flux_J_general}
  \vv J(\vv r, t)
  \equiv
  -D \mynabla \rho(\vv r, t)
  +
  v(\vv r)
  \left[
    \boldsymbol{\mathcal I}_\rho^{(1)}
    \rho(\vv r, t)
    +
    \boldsymbol{\mathcal I}_p^{(1)}
    \vv p(\vv r, t)    
  \right],
\end{equation}
and the matrix flux
\begin{equation}
  \label{eq:flux_M_general}
  \vv M(\vv r, t)
  \equiv
  -D \mynabla \vv p(\vv r, t)
  +
  v(\vv r)
  \left[
    \boldsymbol{\mathcal I}_\rho^{(2)}
    \rho(\vv r, t)
    +
    \boldsymbol{\mathcal I}_p^{(2)}
    \vv p(\vv r, t)    
  \right].
\end{equation}
The quantities
\(
\boldsymbol{\mathcal I}_{\rho,p}^{(1,2)}
\)
account for a possibly restricted acceptance range $\mathcal A$ for propulsion, and are defined as
\begin{equation}
  \begin{split}
    \label{eq:definition_Is_general}
    \boldsymbol{\mathcal I}_\rho^{(1)}
    &\equiv
    \frac{1}{\mathcal S_d}
    \int\limits_{\mathcal A} \df \hn ~ \hn,
    \\
    \boldsymbol{\mathcal I}_p^{(1)}
    =
    d ~\boldsymbol{\mathcal I}_\rho^{(2)}
    &\equiv
    \frac{d}{\mathcal S_d}
    \int\limits_{\mathcal A} \df \hn ~ \hn\hn,
    \\
    \boldsymbol{\mathcal I}_p^{(2)}
    &\equiv
    \frac{d}{\mathcal S_d}
    \int\limits_{\mathcal A} \df \hn ~ \hn\hn\hn.
  \end{split}
\end{equation}
Note that for  orientationally unrestricted propulsion, \ie, $\mathcal A = \mathcal S_d$ and $\chi_{\mathcal A}=1$, the only contributing integrals are
\begin{equation}
\boldsymbol{\mathcal I}_p^{(1)}
=
d ~\boldsymbol{\mathcal I}_\rho^{(2)}
=
\vv 1,
\label{eq:coeffs_active-passive}
\end{equation}
with the unit matrix $\vv 1$.


\subsection{Steady State and Continuity Conditions}
\label{sec:cont-cond}
For the remainder, we focus on the steady-state particle density and polarization. Vanishing time-derivatives in Eqs.~\eqref{eq:mom_equ_rho_general} and \eqref{eq:mom_equ_pol_general} provide the stationarity conditions
\begin{align}
  \label{eq:mom_equ_rho_general_steady_state}
  D\mynabla^2 \rho(\vv r)
  &=
    \mynabla \cdot
    \left[
    v(\vv r)
    \left(
        \boldsymbol{\mathcal I}_\rho^{(1)}
    \rho(\vv r)
    +
    \boldsymbol{\mathcal I}_p^{(1)}
    \vv p(\vv r)    
    \right)
    \right]
  \\[0.5em]
  \nonumber
  D \mynabla^2 \vv p(\vv r)
  &=
    (d-1)\Dr \vv p(\vv r)
  \\
  \label{eq:mom_equ_pol_general_steady_state}  
  &+
    \mynabla \cdot
    \left[
    v(\vv r)
    \left(
    \boldsymbol{\mathcal I}_\rho^{(2)}
    \rho(\vv r)
    +
    \boldsymbol{\mathcal I}_p^{(2)}
    \vv p(\vv r)    
    \right)
    \right].
\end{align}
Furthermore, it turns out that for all setups considered here the no-flux boundary conditions imply that the steady state flux $\vv J(\vv r)$ vanishes at each point in space.
Equation \eqref{eq:flux_J_general} then implies
\begin{equation}
  \label{eq:no_flux_general}
  \mynabla\rho
  =
  \frac{v}{D}
  \left(
    \boldsymbol{\mathcal I}_\rho^{(1)}
    \rho
    +
    \boldsymbol{\mathcal I}_p^{(1)}
    \vv p    
  \right),
\end{equation}
which we substitute into Eq.~\eqref{eq:mom_equ_pol_general_steady_state}.

Let us now consider two domains of constant activity, $\mathcal V_\te{A}$ and $\mathcal V_\te{a}$, whose interface is described by a hyperplane $\mathcal C$, as sketched in Fig.~\ref{fig:setup_2d_general}. For $\vv r \in \mathcal V_{\te{A},\te{a}}$, the Janus particle propels at constant swim speed $v_{\te{A},\te{a}}$, given that its orientation lies within the acceptance range of nudging. Upon crossing the interface $\mathcal C$, the swimmer experiences a sudden change in its activity.  The respective solutions $\rho_{\te{A},\te{a}}(\vv r)$ and $\vv p_{\te{A},\te{a}}(\vv r)$ of the steady-state moment Eqs.~\eqref{eq:mom_equ_rho_general_steady_state} and \eqref{eq:mom_equ_pol_general_steady_state} within each domain $\mathcal V_{\te{A},\te{a}}$ have to be matched at the interface $\mathcal C$. Besides continuity of $\rho$ and $\vv p$ itself, we demand the normal components
\(
\vv J \cdot \vv n_\mathcal C
\)
and
\(
\vv M \cdot \vv n_\mathcal C
\)
of both fluxes   to be continuous at each point $\vv r_\mathcal C$ along the interface $\mathcal C$ \cite{risken}.  The surface normal $\vv n_{\mathcal C}$ is defined to point towards $\mathcal V_\te{a}$.
Computing the limits
\(
\lim_{|\boldsymbol{\epsilon}| \to 0}
r_{\mathcal C} \pm \boldsymbol{\epsilon}
\)
in Eqs.~\eqref{eq:flux_J_general} and \eqref{eq:flux_M_general} delivers the following two continuity conditions
\begin{equation}
  \label{eq:cont_condition_general}
  \vv n_{\mathcal C}
  \cdot
  \left(
    \vv J_\te{a} - \vv J_\te{A}
  \right)
  = 0,
  \qquad
  \vv n_{\mathcal C}
  \cdot
  \left(
    \vv M_\te{a} - \vv M_\te{A}
  \right)
  = 0.  
\end{equation}
While the first of these relations is obvious from $\vv J \equiv \vv 0$, both follow formally  by integrating the (stationary) moment Eqs.~\eqref{eq:mom_equ_rho_general_steady_state} and \eqref{eq:mom_equ_pol_general_steady_state} over an infinitesimal area around some point on the interface $\mathcal C$ and exploiting the divergence theorem.

In the next section, we discuss analytical solutions for the density $\rho$ and polarization $\vv p$ for a quasi-onedimensional system.

\subsection{Activity step in an infinite planar arena}
\label{sec:plan-active-pass}

Consider a situation in which the Janus swimmer faces an orientation-independent activity with a step at an infinite planar interface in $d=2$ dimensions.  
For $x < \xif$, the particle propels at a swim speed $\vA$, which abruptly reduces to $\va < \vA$ upon crossing the interface at $x=\xif$. 
Due to the translational symmetry of the system in the $y$-direction, we project the particle dynamics onto the $x$-axis by replacing $(\vv r, \hn)$ by $(x, \cos\theta)$ in the general equations above. 

In this situation, the only non-zero coefficients~\eqref{eq:definition_Is_general} are given by Eq.~\eqref{eq:coeffs_active-passive} and thus the flux balance \eqref{eq:no_flux_general} reads
\begin{equation}
  \label{eq:flux_balance_act-pass}
  \rho'(x)
  =
  \frac{v(x)}{D} p(x).
\end{equation}
Plugging this relation into the moment Eq.~\eqref{eq:mom_equ_pol_general_steady_state} yields
\begin{equation}
  \label{eq:ode_pol_act-pass}
  p''(x)
  =
  \frac{p(x)}{\lambda^2(x)}
  +
  \frac{v'(x) \rho(x)}{2 D},
\end{equation}
where we defined the natural relaxation length
\begin{equation}
  \label{eq:def_lambda_act-pass}
  \lambda(x)
  \equiv
  \left[
    \frac{\Dr}{D}
    +
    \frac{v^2(x)}{2D^2}
  \right]^{-1/2}.
\end{equation}
For the considered piecewise constant activity profile, Eq.~\eqref{eq:ode_pol_act-pass} boils  down within each halfspace to
\(
p_{\te{A},\te{a}}''(x)
=
p_{\te{A},\te{a}}(x)/\lambda_{\te{A},\te{a}}
\)
with the characteristic length scales $\lambda_{\te{A},\te{a}}$ defined by Eq.~\eqref{eq:def_lambda_act-pass}.
Their relation can be expressed as
\begin{equation}
  \label{eq:lambda_ratio_act-pass}
  \frac{\lA}{\la}
  =
  \sqrt{\frac{1 + \mathcal P_\te{a}}{1 + \mathcal P_\te{A}}},
  \qquad
  \mathcal P_{\te{A},\te{a}}
  \equiv
  \frac{v_{\te{A},\te{a}}^2}{2 D \Dr},
\end{equation}
with the P\'{e}clet numbers $\mathcal P_{\te{A},\te{a}}$ weighing active versus diffusive transport rates in the respective regions.

The general solutions of the polarization profiles are 
\begin{equation}
  \label{eq:pol_gen_sol_act-pass}
  p_{\te{A},\te{a}}(x)
  =
  C_{\te{A},\te{a}}^{(1)} ~ \te{e}^{x/\lambda_{\te{A},\te{a}}}
  +
  C_{\te{A},\te{a}}^{(2)} ~ \te{e}^{-x/\lambda_{\te{A},\te{a}}}
\end{equation}
and the density profile follows by integrating Eq.~\eqref{eq:flux_balance_act-pass} from an arbitrary reference point $x_0$:
\begin{equation}
  \label{eq:rho_gen_sol_act-pass}
  \rho(x)
  =
  \rho(x_0)
  +
  \int\limits_{x_0}^x \df x ~
  \frac{v(x)}{D} p(x).
\end{equation}
The boundary term $\rho(x_0)$ follows from normalization condition for the density and the integration constants $C_{\te{A},\te{a}}^{(1,2)}$ are determined by boundary and matching conditions at the activity step. The latter ones read
\begin{align}
  \label{eq:cont_rho_pol_act-pass}
  \rho_\te{A}(\xif) = \rho_\te{a}(\xif),
  &\qquad
    p_\te{A}(\xif) = p_\te{a}(\xif),
  \\[0.5em]
  \label{eq:jump_cond_pol_act-pass}  
  p_\te{a}'(\xif) - p_\te{A}'(\xif)
  &=
    \frac{\rho_{\te{A},\te{a}}(\xif)}{2D}(\va - \vA),
\end{align}
where the second line 
follows from Eq.~\eqref{eq:cont_condition_general} while using
\(
v'(x)
=
(\va - \vA) \delta(x - \xif)
\), with the delta function $\delta(x)$.


The natural boundary conditions in an infinite arena are \cite{gardinger}
\(
p(|x|\to \infty) = 0.
\)
The continuity condition \eqref{eq:cont_rho_pol_act-pass} then implies that the polarization profiles around an activity step at the origin, $\xif=0$, take the form
\begin{equation}
  \label{eq:pol_profiles_nat_bound}
  p_{\te{A},\te{a}}(x)
  =
  p_\text{max} \te{e}^{-|x|/\lambda_{\te{A},\te{a}}},
\end{equation}
with an unknown maximum polarization $p_\text{max}$.
The density profile follows via Eq.~\eqref{eq:rho_gen_sol_act-pass} as
\begin{align}
  \label{eq:rho_step<}
  \rho_\te{A}(x)
  &=
    \rho_\te{A}
    +
    \frac{p_\text{max}}{D}
    \vA \lA \te{e}^{x/\lA},
  \\[0.5em]
  \label{eq:rho_step>}  
  \rho_\te{a}(x)
  &=
    \rho_\te{A}
    +
    \frac{p_\text{max}}{D}
    \left(
        \vA \lA + \va \la
    -
    \va \la \te{e}^{-x/\la}
    \right),
\end{align}
so that $\rho(x\to \pm\infty)$ attains the regional constant bulk densities $\rho_{\te{a},\te{A}}$. 

A suitable order parameter for the polarization at the interface is the relative maximum polarization $p_\text{max}/\rho(0)$. The condition \eqref{eq:jump_cond_pol_act-pass} allows to express it as
\begin{equation}
  \begin{split}
    \label{eq:pol_peak_nat-bound}
    \frac{p_\text{max}}{\rho(0)}
    &=
    \frac{\vA - \va}{2D}
    \frac{\lA \la}{\lA + \la}
    =
    \frac{1}{\sqrt{2}}
    \frac{\sqrt{\mathcal P_\te{A}} - \sqrt{\mathcal P_\te{a}}
    }{
      \sqrt{1 + \mathcal P_\te{A}} + \sqrt{1 + \mathcal P_\te{a}}
    }.
  \end{split}
\end{equation}
Its sign is via Eq.~\eqref{eq:pol_profiles_nat_bound} shared by the whole polarization profile $p(x)$ and solely determined by the difference in the swim speeds $\vA - \va$. The Janus swimmer thus preferably points from the more into the less active region.
The maximum $1/\sqrt{2}$ of $|p_\te{max}|/\rho(0)$ is reached for $\mathcal P_\te{A} \to \infty$ and $\mathcal P_\te{a}=0$ (or vice versa). It is less than 1, because we consider  particle rotations in two-dimensions in a projection onto the ($x$-)axis of the activity gradient.

By Eqs.~\eqref{eq:pol_profiles_nat_bound}--\eqref{eq:pol_peak_nat-bound},  $p_{\te{A},\te{a}}(x)/\rho(0)$ and $[\rho(x)-\rho_\te{A}]/\rho(0)$ are uniquely determined. The bulk density ratio 
\begin{equation}
  \label{eq:dentiy_ratio_nat-bound}
  \frac{\rho_\te{a}}{\rho_\te{A}}
  =
  \frac{\la}{\lA}
  =
  \sqrt{\frac{1 + \mathcal P_\te{A}}{1 + \mathcal P_\te{a}}}.
\end{equation}
is derived in App.~\ref{sec:dens-ratio-total}. The following section  demonstrates the generality of this result, which also holds for arbitrary activity profiles mediating between the two bulk states.

\begin{figure}[tb!]
  \centering
  \includegraphics[width=\columnwidth]{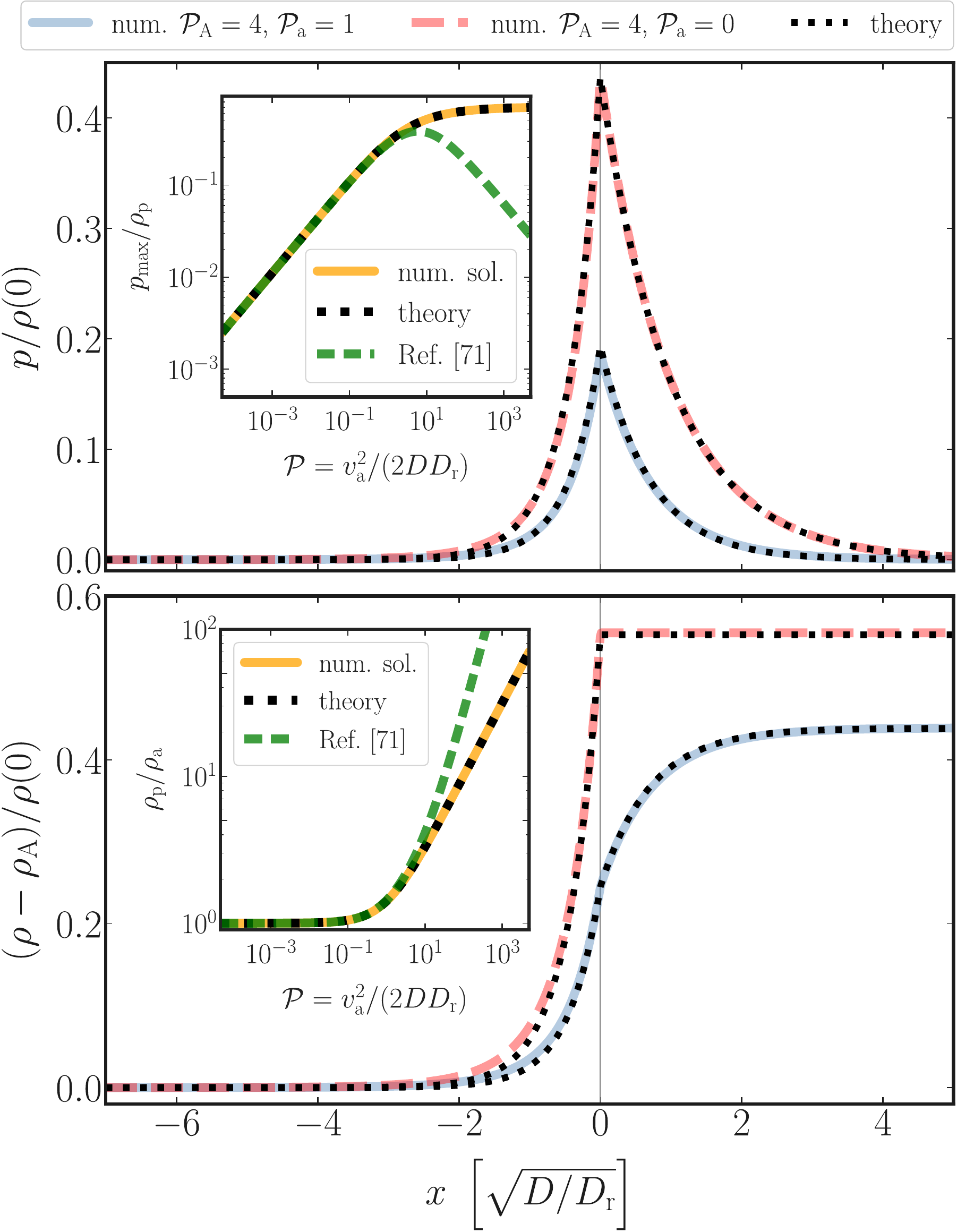}

  \caption{Particle density and polarization profiles --- approximate theory versus exact numerical solutions at an activity step at $x=0$, between a high-activity (A) and a low-activity (a) or entirely passive (p) region.     \emph{Upper panel:} The reduced polarization from Eqs.~\eqref{eq:pol_profiles_nat_bound}, \eqref{eq:pol_peak_nat-bound} closely matches the exact numerical solutions (dashed and solid lines). \emph{Inset:}  the polarization peak at the active-passive interface as function of the active P\'{e}clet number $\mathcal P$  from Eq.~\eqref{eq:pol_peak_nat-bound} (dotted) precisely follows the exact numerical solution (solid line), improving upon earlier predictions~\cite{Fischer2020Quorum-sensingMotility,fischer2020erratum} (dashed). \emph{Lower panel:} The reduced denisty ratio $(\rho-\RA)/\rho(0)$ from Eqs.~\eqref{eq:rho_step<}-\eqref{eq:pol_peak_nat-bound} (dotted) closely matches the exact numerical solutions (dashed and solid lines) \emph{Inset:} the density ratio $\Rp/\Ra$ for the active-passive interface  as a function of the P\'{e}clet number $\mathcal P$ from Eq.~\eqref{eq:dentiy_ratio_nat-bound} (dotted line) precisely follows the numerical results (solid line), improving upon earlier predictions~\cite{Fischer2020Quorum-sensingMotility,fischer2020erratum} (dashed).}
  \label{fig:pol_rho_density_ratio_nat-bound}
\end{figure}

Figure~\ref{fig:pol_rho_density_ratio_nat-bound} compares the approximate theory profiles following from Eqs.~\eqref{eq:pol_profiles_nat_bound}-\eqref{eq:dentiy_ratio_nat-bound} to the (exact) numerical solutions obtained  from Eq.~\eqref{eq:FPE_full} using the method of Ref.~\cite{holubec2018}.
The two cases $0<\va<\vA$ and $0=\va < \vA$ are considered.
The characteristic length scales $\lambda_i$, bulk densities $\rho_i$, and P\'{e}clet numbers $\mathcal P_i$ corresponding to the highly active ($x<0$) and less active/completely passive regions ($x \geq 0$) are distinguished by the subscripts $\te{(A)}$, $\te{(a)}$ and $\te{(p)}$, respectively.
The analytic solutions for the polarization and density profiles in the main panels  nicely agree with the numerical solutions, with slight deviations appearing only for substantial activity.

Such impressive performance of the approximate model is unexpected. Intuitively, the approximation \eqref{eq:moment_expansion_general} should break down for $\mathcal P \simeq 1$. Its remarkable accuracy far beyond this limit becomes more plausible by observing that its moment equations \eqref{eq:flux_balance_act-pass} and \eqref{eq:ode_pol_act-pass} map onto an exactly solvable 2-species model (one-dimensional run-and-tumble \cite{Schnitzer1993TheoryChemotaxis,Tailleur2008StatisticalBacteria,Tailleur2009SedimentationBacteria} accompanied by diffusion \cite{Malakar2018RunAndTumble1D,razin2020entropyProdABPinBox}; see Sec.~\ref{sec:2-species-model}). The latter robustly captures the pertinent physics (though not all quantitative details) at the motility step, for arbitrary P\'{e}clet numbers.
Indeed, the quantitative agreement between approximate analytical and the exact numerical profiles, which we obtain using the method described in Ref.~\cite{holubec2018}, remains very good up to P\'{e}clet numbers on the order of 100, as independently discovered in Ref.~\cite{rodenburg2020_thesis}.  For even higher activities, higher moments in the moment expansion \eqref{eq:moment_expansion_general} become non-zero. However, formally pushing the moment expansion up to the third order actually leads to less precise numerical predictions than our physically motivated second-order approximation, already for intermediate P\'{e}clet numbers on the order of 5.

From the upper panel in Fig.~\ref{fig:pol_rho_density_ratio_nat-bound}, we infer that the relative polarization $p_\te{max}/\rho(0)$ peaks exactly at the motility step with a magnitude given by Eq.~\eqref{eq:pol_peak_nat-bound}, which improves the predictions of Refs.~\cite{Fischer2020Quorum-sensingMotility,fischer2020erratum} (see inset).
The polarization decays exponentially as a function of the distance from the interface.  The polarization layers extend over the characteristic length scale $\lambda$, defined in Eq.~\eqref{eq:def_lambda_act-pass}. Its limiting forms for vanishing and diverging self-propulsion are
\(
\lambda \sim \sqrt{D/\Dr}
\)
, \(
\lambda \sim \sqrt{2}D / v(x)
\), respectively;
i.e.\ the boundary layer is compressed by the particle's self-propulsion. (For a physical interpretation see the discussion section \ref{sec:disc-bound-layer}.)
The inset provides strong numerical evidence that Eq.~\eqref{eq:pol_peak_nat-bound} for the reduced polarization peak generally holds for arbitrarily large activity steps.
The lower panel of Fig.~\ref{fig:pol_rho_density_ratio_nat-bound} presents the corresponding density profiles. For the active-passive interface, the density profile remains constant at the bulk density $\rho = \Rp$ throughout the whole passive region by virtue of Eq.~\eqref{eq:flux_balance_act-pass}. On crossing the interface, it decays to the bulk density $\rho_\te{A} < \Rp$ pertaining to the active region over a length scale $\lambda_\te{A}$. In the case of finite activity in both regions $(0 < \va < \vA)$, the density still displays a kink at the interface, but decays over the length scales $\lambda_\te{a,A}>0$ towards its bulk density values $\rho_\te{a,A}$.
The inset of the lower panel of Fig.~\ref{fig:pol_rho_density_ratio_nat-bound} compares the analytical expression \eqref{eq:dentiy_ratio_nat-bound} for the bulk density ratio $\Rp/\Ra$ at an active-passive interface with the exact numerical solution. Over a vast range of P\'{e}clet numbers $\mathcal P$, both curves perfectly overlap, demonstrating the robustness of Eq.~\eqref{eq:dentiy_ratio_nat-bound}, and again improving predictions by Refs.~\cite{Fischer2020Quorum-sensingMotility,fischer2020erratum}.



The following section points out a number of interesting consequences of the above findings and puts them into a broader context.

\subsection{Various consequences and ramifications: wall accumulation, MIPS, swim pressure}
\label{sec:relation-many-body}

\paragraph*{Wall Accumulation:}
\label{sec:wall-accumulation}

Equations.~\eqref{eq:flux_balance_act-pass} and \eqref{eq:ode_pol_act-pass} can serve to calculate (approximate) polarization and density profiles at various kinds of interfaces or boundaries, which enter the governing equations only via different boundary conditions. As a concrete example, consider  an active particle confined by hard reflecting  walls~\cite{risken}. Our model captures the well-known effects of polarization towards and accumulation at walls \cite{hermann2020PolStateFct,Speck2016IdealBulkPressure,Wagner2017ABPsUnderConfinement,Elgeti2013WallAccumulation,Malakar2018RunAndTumble1D,razin2020entropyProdABPinBox}.
The simplest way to compute specific profiles is by exploiting the equivalence between the approximate model, Eqs.~\eqref{eq:flux_balance_act-pass} and  \eqref{eq:ode_pol_act-pass}, and the 2-species model (see Sec.~\ref{sec:2-species-model}), and compute $\rho(x)$ and $p(x)$ within the latter, as done in \cite{Malakar2018RunAndTumble1D,razin2020entropyProdABPinBox}.
The polarization and density layers near the wall are still determined by the same boundary-layer width $\lambda$ in Eq.~\eqref{eq:def_lambda_act-pass} and exhibit the same physics as those at motility steps (see the discussion in Sec.~\ref{sec:discussion}).

\paragraph*{MIPS:}
\label{sec:mips}
An extensively studied feature of active-Brownian-particle suspensions is so-called motility-induced phase separation (MIPS) \cite{Cates2015MIPS,Cates2013WhenSeparation,paliwal2018ChemPot,hermann2019MIPS,Solon2018GeneralizedMatter,solon2018GenTDofMIPS,Patch2018CurvDepTension}. In a nutshell, it relies on two ingredients: (i) the slowing down of particles in ``crowded'' areas \cite{Schnitzer1993TheoryChemotaxis}, and (ii) active-particle accumulation in low-motility areas \cite{Cates2015MIPS}.
Our results for the polarization and density profiles of a single (overdamped) Janus swimmer can, at least on a qualitative level, closely mimic the effects observed for MIPS if applied to a gas of non-interacting active Janus spheres experiencing an inhomogeneous activity profile.
As also pointed out in Ref.~\cite{rodenburg2020_thesis}, such a gas exhibits the following features observed in MIPS:
\begin{enumerate}[(i)]
\item The system divides into two bulk areas with different bulk densities. Particles move slower (faster) in the denser (less dense) regions \cite{Cates2015MIPS,solon2018GenTDofMIPS} (cf.~Eq.~\eqref{eq:dentiy_ratio_nat-bound} and lower panel of Fig.~\ref{fig:pol_rho_density_ratio_nat-bound}).
\item An interfacial region forms between the two bulk areas, where particles, on average,  point into the denser phase \cite{hermann2019MIPS,solon2018GenTDofMIPS,paliwal2018ChemPot,omar2020swimPressure} (cf.~upper panel of Fig.~\ref{fig:pol_rho_density_ratio_nat-bound}).
  Note that for active Lennard-Jones particles, an opposite polarization was observed \cite{paliwal2017SurfTensionLJSys}.
  \item Smaller activity  corresponds to a wider  polarization layer \cite{Patch2018CurvDepTension} (cf.~Eq.~\eqref{eq:def_lambda_act-pass}).
\end{enumerate}

In the particular case of quorum-sensing particles, adjusting their activity according to the local density (e.g.\ via chemical signaling) \cite{Miller_2001_Quorum},
Eq.~\eqref{eq:dentiy_ratio_nat-bound} for the bulk density ratio can be used to improve corresponding literature results~\cite{Fischer2020Quorum-sensingMotility,fischer2020erratum} based on a dynamic mean-field theory.

\paragraph*{Swim pressure/total polarization:}
\label{sec:swim-press-polar}

In suspensions of active Brownian particles a ``swim pressure'' \cite{Solon2015PressureSpheres,hermann2019MIPS} emerges due to local polarization of the swimmers in the vicinity of various interfaces (sedimentation \cite{hermann2018Sediment,vachier2019SedimentABP,enculescu2011PolarOrderGrav,ginot2018SedimentatPolPress}, walls \cite{hermann2020PolStateFct,Speck2016IdealBulkPressure,Wagner2017ABPsUnderConfinement,Elgeti2013WallAccumulation}, MIPS \cite{hermann2020PolStateFct,paliwal2018ChemPot,hermann2019MIPS,Solon2018GeneralizedMatter,solon2018GenTDofMIPS,prymidis2016VapLiquCoeyInLJSys,paliwal2017SurfTensionLJSys}).
The swim pressure is, up to a prefactor, determined by the total polarization $P_\te{tot}$, which is defined as the integrated local polarization profile $p(x)$ over some (sub-)volume \cite{hermann2020PolStateFct},  and thus also points from the more active to the less active region.
For the motility step considered in the previous section, Eqs.~\eqref{eq:pol_profiles_nat_bound} and \eqref{eq:dentiy_ratio_nat-bound} imply that within our approximate model the total polarization reads
\begin{equation}
  \begin{split}
      \label{eq:Pol_total}
      P_\te{tot}
      =
      \int_{-\infty}^\infty \df x ~ p(x)
      &=
      p_\te{max} (\la + \lA)
      =
      \frac{\vA\rho_\te{A} - \va\rho_\te{a}}{2 \Dr}.
  \end{split}
\end{equation}
A derivation of the last equality is given in  App.~\ref{sec:dens-ratio-total}. The total polarization is thus determined by the difference of the bulk fluxes $v_{\te{A},\te{a}} \rho_{\te{A},\te{a}}$ in the bulk regions and the rotational diffusion coefficient $\Dr$. This agrees with the general result of Hermann and Schmidt \cite{hermann2020PolStateFct} and verifies that the total polarization induced by an activity step is a state function. There does not seem to be any straightforward analogy with equilibrium phase coexistence, similar as shown in \cite{caprini2020velocityAlign} for the velocity alignment of active Brownian particles.

\subsection{Generalizations}
We now show that Eq.~\eqref{eq:dentiy_ratio_nat-bound} for the bulk density ratio actually holds for the full ABP model~\eqref{eq:FPE_full} and arbitrary onedimensional activity variations, and we extend the above discussion to finite domains.

\paragraph*{Density ratio:}
\label{sec:density-ratio}
Consider an arbitrary activity profile $v(x)$ that mediates between two bulk regions of respectively constant activity.
Using Eq.~\eqref{eq:FPE_full},
the corresponding stationary FPE for the probability density $f(x,\theta)$ reads
\begin{equation}
  \label{eq:FPE_1D}
  0
  =
  - \partial_x \mathcal J
  - \partial_\theta \mathcal J_\theta
  =
  D \partial_x^2 f
  +
  \Dr \partial_\theta^2 f
  -
  \partial_x
  \left(
    v \cos\theta f
  \right),
\end{equation}
with the (angle-resolved) translational and rotational currents
\begin{align}
  \label{eq:J_trans_full}
  \mathcal J(x,\theta)
  &\equiv
    -D \partial_x f(x,\theta)
    +
    v(x) \cos\theta f(x,\theta),
  \\
  \label{eq:J_rot_full}
  \mathcal J_\theta(x,\theta)
  &\equiv
    -\Dr \partial_\theta f(x,\theta).
\end{align}
The distribution function $f$ is expanded into the Fourier series
\begin{equation}
  \label{eq:moment_expansion_full}
  f(x,\theta)
  =
  \frac{\rho(x)}{2\pi}
  +
  \frac{1}{\pi}
  \sum\limits_{n=1}^{\infty} f_n(x) \cos(n\theta),
\end{equation}
with coefficients $f_n = \langle \cos(n\theta) \rangle = \int_0^{2\pi} d\theta\, \cos(n\theta) f(x,\theta)$. In accord with our previous discussion, the zeroth and the first coefficients are given by the density $\rho \equiv \langle 1 \rangle$ and the polarization $p \equiv f_1 = \langle \cos\theta \rangle$, respectively.
The orientation-averaged flux thus reads
\begin{equation}
  \label{eq:flux_1d_general}
  J (x)
  \equiv
  \langle \mathcal J(x,\theta) \rangle  
  =
  -D \rho'(x)
  +
  v(x) p(x).
\end{equation}
Similarly, multiplication of the FPE \eqref{eq:FPE_1D} by $\cos\theta$ and integration over $\theta$ yields the differential equation for the polarization
\begin{equation}
  \label{eq:FPE_1D_sec_mom}
  0
  =
  D p''
  -
  \Dr p
  -
  \frac{v}{2}\rho'
  -
  \frac{\rho}{2}v'  
  -
  \frac12
  \partial_x
  \left(
    v f_2
  \right),
\end{equation}
where we used the identity
\(
2 \cos^2\theta
=
1 + \cos(2\theta)
\).
Note that the last term in Eq.~\eqref{eq:FPE_1D_sec_mom} was absent in the previous discussions due to the applied closure relation
\(
f_2
=
\langle
\cos(2\theta)
\rangle
=
0
\).
Isolating $p$ from Eq.~\eqref{eq:FPE_1D_sec_mom} and plugging it into Eq.~\eqref{eq:flux_1d_general} yields
\begin{equation}
  \label{eq:flux_Deff_general}
  J(x)
  =
  -\Deff(x) \rho'(x)
  -
  \frac12 \rho \Deff'
  +
  J_{v'}(x),
\end{equation}
with the position-dependent effective diffusivity
\begin{equation}
  \label{eq:Deff_def}
  \Deff(x)
  \equiv
  D + \frac{v^2(x)}{2\Dr}
\end{equation}
and the flux
\begin{equation}
  \label{eq:J_lambda}
  J_{v'}
  \equiv
  \frac{D}{\Dr} v(x) p''(x)
  +
  \frac{v}{2\Dr}
  \partial_x
  \left(
    v f_2
  \right).
\end{equation}
Equation~\eqref{eq:flux_Deff_general} constitutes a generalized version of Fick's law.  The first contribution, $-\Deff \rho'$, accounts for isotropic diffusive transport. The effective diffusivity $\Deff \geq D$ is enhanced by the swimmer's short-term ballistic motion.  The second term, $\rho \Deff'/2$, accounts for the spatial dependency of this effective diffusivity, and its prefactor $1/2$ for the directionality of the active velocity.
The last contribution, $J_{v'}$, represents the influence of the polarization $p(x)$ and higher moments $f_n$, $n>1$ on the local density.

The condition~\eqref{eq:flux_Deff_general} of a vanishing steady-state flux, $J(x) = 0$, yields
\begin{equation}
  \label{eq:ode_rho_Deff}
  \frac{\rho'}{\rho}
  =
  - \frac12
  \frac{\Deff'}{\Deff}
  +
  \frac{J_{v'}}{\rho \Deff}.  
\end{equation}
Integrating this equation from a reference point $x_0$ up to an arbitrary position $x$, we find
\begin{equation}
  \label{eq:formal_solution_density_ratio}
  \frac{\rho(x)}{\rho(x_0)}
  =
  \sqrt{ \frac{\Deff(x_0)}{\Deff(x)} } ~
  \te{exp}
  \left\{
    \mathcal U [v](x_0,x)
  \right\},  
\end{equation}
where we introduced the functional
\begin{equation}
  \label{eq:density_ratio_functional}
  \mathcal U[v](x_0,x)
  \equiv
  \int\limits_{x_0}^x \df \tilde x ~
  \frac{
    J_{v'}(\tilde x)
  }{
    \rho(\tilde x) \Deff(\tilde x)}.
\end{equation}
Equation \eqref{eq:formal_solution_density_ratio} shows that density ratios are determined by the ratio of the corresponding effective diffusion coefficients, corrected by the exponential of a complicated functional $\mathcal U[v](x_0,x)$
of the $J_{v'}/\Deff$ and thus the activity profile $v(x)$. It is therefore not suitable for a computation of the full density profile. However, if the integration bounds $x_0$ and $x$ in \eqref{eq:formal_solution_density_ratio} and \eqref{eq:density_ratio_functional} are sufficiently far away from the activity variations, such that $\rho(x)$ and $\rho(x_0)$ correspond to the bulk densities, one can proof that the functional vanishes, $\mathcal U[v](x_0,x) = 0$, irrespective of the activity profile $v(x)$ (see App.~\ref{sec:vanishing-integral}).  We therefore find that bulk density ratio is generally given by
\begin{equation}
  \label{eq:density_ratio_general}
  \frac{\rho(x)}{\rho(x_0)}
  =
  \sqrt{ \frac{\Deff(x_0)}{\Deff(x)} }\,.
\end{equation}
By the definition of the P\'{e}clet number in Eq.~\eqref{eq:lambda_ratio_act-pass}, this result is seen to coincide with Eq.~\eqref{eq:dentiy_ratio_nat-bound}.
For highly persistent swimmers, \ie, $v^2(x) \gg 2 D \Dr$, Eq.~\eqref{eq:density_ratio_general} reduces to the well-known relation \cite{Schnitzer1993TheoryChemotaxis,Cates2012DiffusivePhysics}
\(
\rho(x)/\rho(x_0)
=
v(x_0)/v(x).
\)
To conclude, the bulk density ratio $\rho(x)/\rho(x_0)$ is generally independent of the exact shape and magnitude of the activity variations.  As a consequence,  on a mesoscopic scale, at which all moments $f_{n>0}$ in Eq.~\eqref{eq:moment_expansion_full} are negligibly small compared to $\rho$, the coarse-grained version of the flux balance~\eqref{eq:flux_Deff_general} reads
\(
\overline{J}(x)
=
-\Deff(x) \rho'(x)
-
\frac12 \rho \Deff'
\).

\begin{figure}[tb!]
  \centering
 \includegraphics[width=\columnwidth]{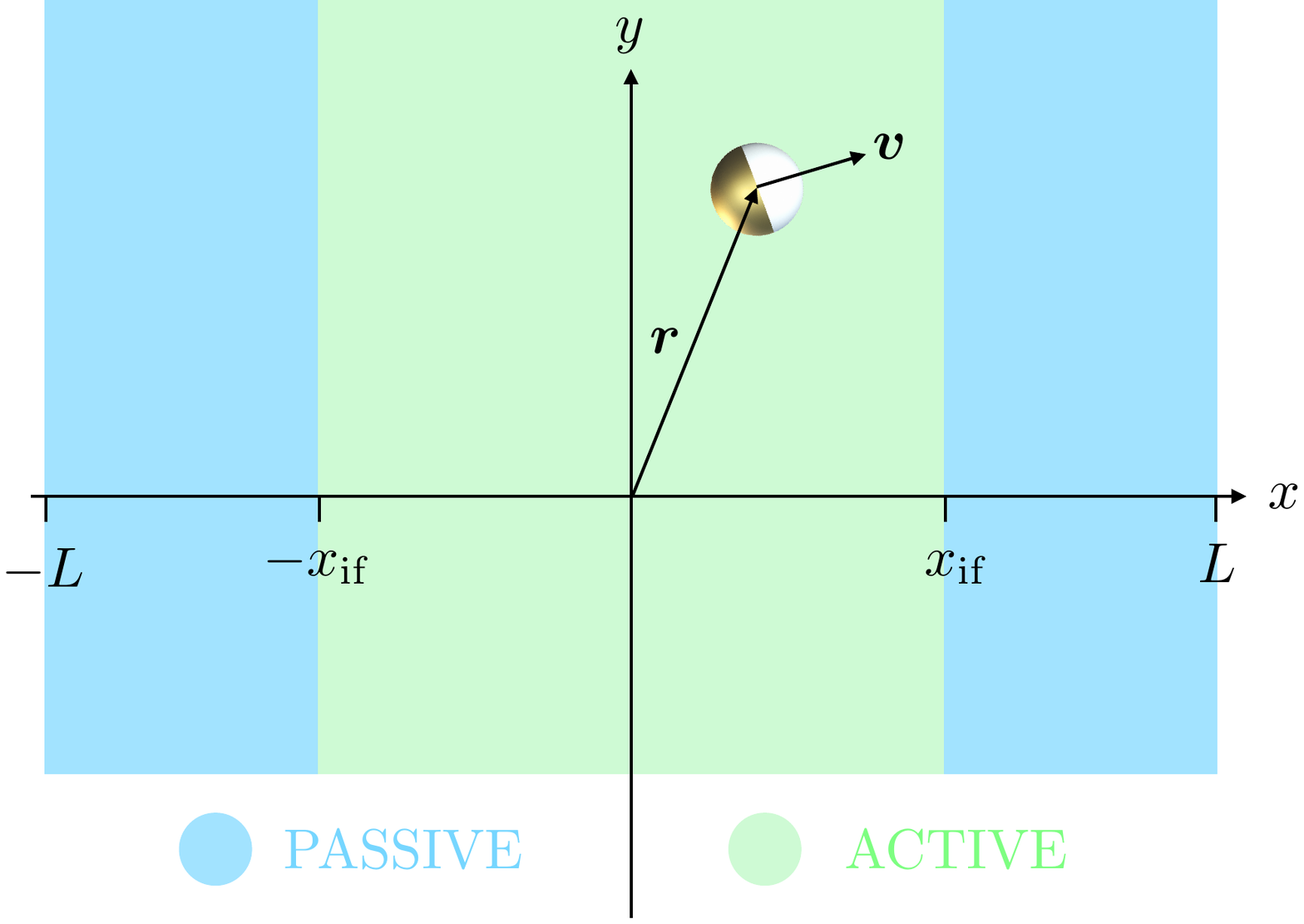}
  \caption{Sketch of a Janus particle with orientation $\vv n =(\cos\theta, \sin\theta)$ and position $\vv r=(x,y)$. It propels actively with velocity $\bm v = \va \hn$ along its symmetry axis as long as $|x|<\xif$.  Otherwise it undergoes ordinary translational and rotational diffusion.  The system has periodic boundaries located at $\pm L$.}
  \label{fig:setup_theory_finite-system}
\end{figure}

\paragraph*{Finite domains:}
\label{sec:symmetrin-case}
The experiment described in the companion article \cite{Soeker2020ActivityFieldsPRL} was performed in a finite arena. We therefore next derive the appropriate analytic solutions $\rho(x)$ and $p(x)$ for an activity step in a rectangular domain of length $2L$, comprising a central active region of length $2 \xif$ and two adjacent passive regions interconnected by periodic boundaries, as sketched in Fig.~\ref{fig:setup_theory_finite-system}. The corresponding activity profile is given by
\(
v(x)
=
\va \Theta(\xif-|x|).
\)
The symmetry of this setup allows us to consider only the positive half space with active-passive interface at $x=\xif$.
The respective polarization and density profiles in the active and passive regions are still given by Eqs.~\eqref{eq:pol_gen_sol_act-pass} and \eqref{eq:rho_gen_sol_act-pass}.
By virtue of the system's symmetry and the imposed periodic boundary conditions at $x=\pm L$, the polarization must obey $\Pa(0)=0$ and $\Pp(L) = 0$. Hence, the polarization profiles in the active and passive regions read
\begin{equation}
  \label{eq:pol_act_pass_general_finite}
  \Pa(x)
  =
  \Ca
  \sinh
  \left(
    \frac{x}{\la}
  \right),
  \quad
  \Pp(x)
  =
  \Cp
  \sinh
  \left(
    \frac{L-x}{\lp}
  \right).
\end{equation}
Taking $x_0=0$ as reference point in Eq.~\eqref{eq:rho_gen_sol_act-pass}, the density profile on the active side ($0 \leq x < \xif$) reads
 \begin{equation}
   \label{eq:rho_solution_active_finite}
   \Ra(x)
   =
   \Ra(0)
   +
   \frac{\Ca \va \la}{D}
   \left[
     \cosh
     \left(
       \frac{x}{\la}
     \right)
     -
     1
   \right].
 \end{equation}
The corresponding density profile on the passive side ($\xif \leq x \leq L$) is constant,
 \(
\Rp(x) \equiv \Ra(0).
 \)
The integration constants $\Ca$ and $\Cp$ and the reference density $\rho_{\rm a}(0)$ are uniquely determined by the continuity conditions
\begin{align}
  \label{eq:matching_cond_1}
  \Pp(\xif) &= \Pa(\xif),
  \\
  \label{eq:matching_cond_2}  
  \Pa'(\xif) - \Pp'(\xif)
         &=
           \frac{\va}{2D}
           \rho(\xif),
\end{align}
and the normalization condition on $\rho(x)$.
The detailed calculation can be found in App.~\ref{sec:integration_constants}.

\begin{figure}[tb!]
  \centering
  \includegraphics[width=\columnwidth]{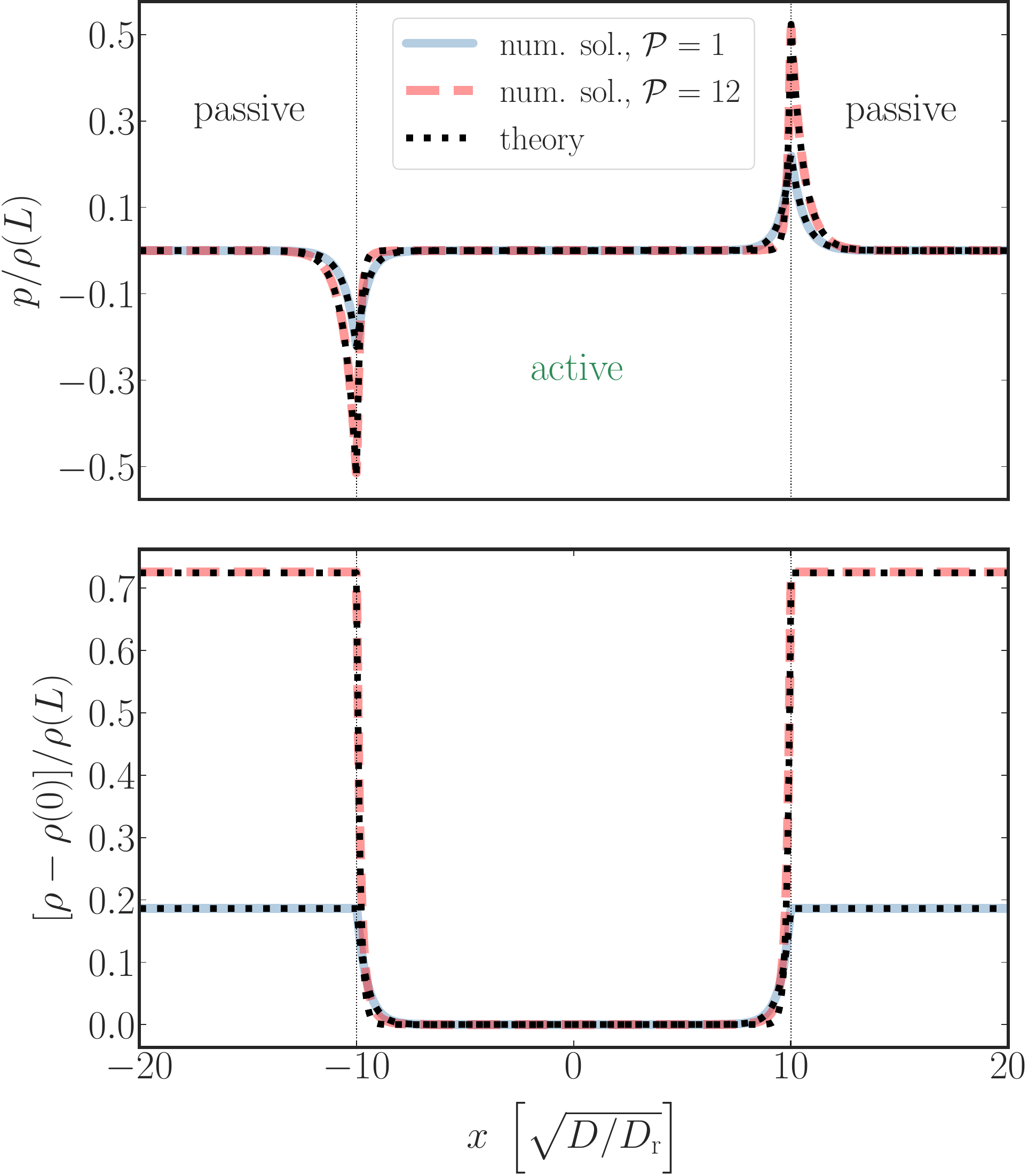}
  \caption{Polarization-density patterns for a finite active domain bounded at $x=\pm\SI{10}{\lp}$ by two passive margins extending to  $\pm\SI{20}{\lp}$. Our approximate theory \eqref{eq:pol_compact_act}-\eqref{eq:rho_compact_pass} for the particle density and polarization (dotted) fares well compared to the exact numerical solutions (dashed and solid lines).}
  \label{fig:pol_rho_density_ratio_finite-size}
\end{figure}

Let us assume that the active and passive regions are sufficiently large compared to the decay lengths $\lambda_{\te{A},\te{a}}$ to maintain a scale separation between boundaries and bulk. Then, the polarization and density profiles can be written as
\begin{align}
  \label{eq:pol_compact_act}
  \Pa(x)
  &=
    \frac{1}{2L}
    \frac{
    P_{\te{max}}
    }{
    1-(1-r_\rho)\frac{\xif-\la}{L}
    }
    \frac{\sinh(x/\la)}{\sinh(\xif/\la)},
  \\[0.5em]
  \Pp(x)
  &=
    \label{eq:pol_compact_pass}
    \frac{1}{2L}
    \frac{
    P_{\te{max}}
    }{
    1-(1-r_\rho)\frac{\xif-\la}{L}
    }
    \frac{\sinh((L-x)/\lp)}{\sinh((L-\xif)/\lp)},
  \\[0.5em]
  \label{eq:rho_compact_act}
  \rho_{\te a}(x)
  &=
    \frac{1}{2L}
    \left[
    \frac{
    1}{
    1-(1-r_\rho)\frac{\xif-\la}{L}
    }
    \right.
  \\
  \nonumber
  &\hspace{0.75cm}+
  \left.
    \frac{
    1-r_\rho}{
    1-(1-r_\rho)\frac{\xif-\la}{L}
    }
    \left(
    \frac{\cosh(x/\la)}{\sinh(\xif/\la)} - 1
    \right)
    \right],
  \\[0.5em]
  \label{eq:rho_compact_pass}
  \rho_{\te p}(x)
  &\equiv
    \rho_{\te a}(\xif)
    =
    \frac{1}{2L}
    \frac{
    1}{
    1-(1-r_\rho)\frac{\xif-\la}{L}
    },
\end{align}
where we employed the short-hand notation
\begin{align}
  \label{eq:Pmax}
  P_{\te{max}}
  &\equiv
  \frac{\va}{2D}
  \frac{\la \lp}{\la + \lp}
  =
  \frac{1}{\sqrt{2}}
  \frac{\sqrt{\mathcal P}
  }{
    1 + \sqrt{1 + \mathcal P}
  },
  \\
  \label{eq:r_rho}
  r_\rho
  &\equiv
  \frac{\la}{\lp}
  =
  \frac{1}{\sqrt{1 + \mathcal P}},
\end{align}
for the maximum (relative) polarization and density ratio, respectively. In Fig.~\ref{fig:pol_rho_density_ratio_finite-size}, we show that these results agree nicely with the exact numerical solutions. In the companion paper \cite{Soeker2020ActivityFieldsPRL}, we further show that they also nicely describe the experimental data.

Note that the total polarization of the system in Fig.~\ref{fig:pol_rho_density_ratio_finite-size},
\(
P_\te{tot}
=
\int_{-L}^L \df x ~ p(x)
\),
vanishes by virtue of the anti-symmetry between the polarization profiles on the left and right activity step. This is in agreement with Ref.~\cite{hermann2020PolStateFct}, stating that $P_\te{tot} = 0$ for systems with vanishing fluxes at their boundary. This means that inhomogeneous activity profiles in closed systems merely operate as local spatial sorting mechanisms for particles of different polarisation.

\subsection{Nudging}
\label{sec:photon-nudging}
To mimic the technique of photon nudging \cite{Qian2013HarnessingNudging,Bregulla2014StochasticNudging,Selmke2018TheoryTransport,Selmke2018TheoryConfinement} exploited by the experimental setup described in the companion article \cite{Soeker2020ActivityFieldsPRL}  (see also Fig.~\ref{fig:setup_exp}), we need to allow the  activity to also depend on the particle orientation. In the nudging regions, the ``fuel'' for the particle's autonomous propulsion is restricted not only spatially but additionally also by an acceptance range of particle orientations.
We now discuss this additional complication.
The angular dependence of the activity profile is modelled as $v(x,\theta) = v(x)\Theta(\alpha-|\theta|)$, where $\alpha$ represents the acceptance angle.  Then, the no-flux condition \eqref{eq:no_flux_general} and the moment from Eq.~\eqref{eq:mom_equ_pol_general_steady_state}
take the modified forms
\begin{align}
  \label{eq:no_flux_nudging}
  \rho'
  &=
    \frac{v}{D}
    \left(
    \mathcal I_\rho^{(1)}\rho
    +
    \mathcal I_p^{(1)}p    
    \right),
  \\[0.5em]
  \label{eq:ode_pol_nudging}
  p''
  &=
    \frac{\Dr}{D} p
    +
    \frac{v}{D}
    \left(
    \mathcal I_\rho^{(2)} \rho'
    +
    \mathcal I_p^{(2)} p'  
    \right)
    +
    \frac{v'}{D}
    \left(
    \rho \mathcal I_\rho^{(2)}
    +
    p \mathcal I_p^{(2)} 
    \right).
\end{align}
According to  the definitions \eqref{eq:definition_Is_general}, the constants $\mathcal I_{\rho,p}^{(1,2)}$, which represent the influence of the restricted acceptance angle $\alpha$, read
\begin{align}
  \label{eq:I_rho_1}
  \mathcal I_\rho^{(1)}
  &=
    \frac{1}{2\pi}
    \int\limits_{-\alpha}^{\alpha} \df \theta
    ~ \cos\theta
    =
    \frac{\sin\alpha}{\pi},
  \\[0.5em]
  \label{eq:I_p_1}
  \mathcal I_p^{(1)}
  &=
    \frac{1}{\pi}
    \int\limits_{-\alpha}^{\alpha} \df \theta
    ~ \cos^2\theta
    =
    \frac{\alpha}{\pi}
    +
    \frac{\sin(2\alpha)}{2\pi},
  \\[0.5em]
  \label{eq:I_rho_2}  
  \mathcal I_\rho^{(2)}
  &=
    \frac{\mathcal I_p^{(1)}}{2}
    =
    \frac{\alpha}{2\pi}
    +
    \frac{\sin(2\alpha)}{4\pi},
  \\[0.5em]
  \label{eq:I_p_2}  
  \mathcal I_p^{(2)}
  &=
    \frac{1}{\pi}
    \int\limits_{-\alpha}^{\alpha} \df \theta
    ~ \cos^3\theta
    =
    \frac{9 \sin\alpha + \sin(3\alpha)}{6\pi}.
\end{align}
The swimmer is nudged to the right if $\alpha < \pi$ and to the left by formally replacing $v \to -v$. The case $\alpha=\pi$ corresponds to no nudging (orientation-independent activity). 
We let the sudden activity step again be located at $x=\xif \equiv 0$ and assume the particle is nudged (``n'') to the right for $x \leq 0$. Upon crossing the interface to $x>0$, it enters a fully active (a) or passive region (p), where its activity does not depend on its orientation.
Within each region, the swim speed $v_i \geq 0$, $i \in \{\te n, \te a, \te p\}$, is constant. 

Plugging the steady-state condition \eqref{eq:no_flux_nudging} into the moment Eq.~\eqref{eq:ode_pol_nudging} yields an equation of the form
\begin{equation}
  \label{eq:matrix_form}
  \vv X'
  =
  \boldsymbol{\Lambda} \vv X,
\end{equation}
where 
\begin{align}
  \vv X
  &\equiv
    (p',p,\rho)^\top,
  \\[0.5em]
  \label{eq:X_Lambda_Definition}
  \boldsymbol{\Lambda}
  &\equiv
  \begin{pmatrix}
    \frac{v_i}{D} \mathcal I_p^{(2)}
    &
    \frac{\Dr}{D}
    +
    \frac{v_i^2}{D^2}
    \mathcal I_\rho^{(2)} \mathcal I_p^{(1)}
    &
    \frac{v_i^2}{D^2}
    \mathcal I_\rho^{(1)} \mathcal I_\rho^{(2)}
    \\[0.2em]
    1 & 0 & 0
    \\[0.2em]
    0
    &
    \frac{v_i}{D} \mathcal I_p^{(1)}
    &
    \frac{v_i}{D} \mathcal I_\rho^{(1)}
  \end{pmatrix}.
\end{align}
For $\alpha < \pi$, all the integrals $\mathcal I_{\rho,p}^{(1,2)}$ from Eqs.~\eqref{eq:I_rho_1}-\eqref{eq:I_p_2} give nonzero contributions. In App.~\ref{sec:continuois-model} we explicitly calculate the eigenvalues $\lambda_{\te{n}_i}^{-1}$ of the matrix $\boldsymbol{\Lambda}$. All of them are real and mutually distinct. 
The general solution to Eq.~\eqref{eq:matrix_form} thus has the structure
\begin{equation}
  \label{eq:general_solution_nudging}
  \vv X
  =
  \sum\limits_{k=1}^3
  C_i \vv w_i \te{e}^{\lambda_{\te{n}_i}^{-1} x},
\end{equation}
where $\vv w_i$ denotes the eigenvector pertaining to the eigenvalue $\lambda_{\te{n}_i}^{-1}$.  The coefficients $C_i$ are determined by boundary and matching conditions.
The intuitive relations 
\begin{equation}
  \label{eq:matching_obvious}
  \rho_\te{n}(0) = \rho_\te{a}(0),
  \qquad
  p_\te{n}(0) = p_\te{a}(0),
\end{equation}
at a nudging-active interface are complemented by the matching condition
\begin{align}
  \label{eq:jump_active_nudging}
  p_\te{a}'(0)-p_\te{n}'(0)
  &=
    \frac{
    \va/2
    -
    \vn\mathcal I_{\rho_\te{n}}^{(2)}}{
    D}
    \rho(0)
    -
    \frac{
    \vn\mathcal I_{p_\te{n}}^{(2)}}{
    D}
    p(0).
\end{align}
It follows from Eq.~\eqref{eq:cont_condition_general} while using that $I_{\rho_\te{a}}^{(2)} = 1/2$ and $I_{p_\te{a}}^{(2)} = 0$ within the active region (see Eq.~\eqref{eq:coeffs_active-passive} and sentence above).  The matching condition for a nudging-passive interface is included as the case $\va = 0$.

\begin{figure}[tb!]
  \centering
\includegraphics[width=\columnwidth]{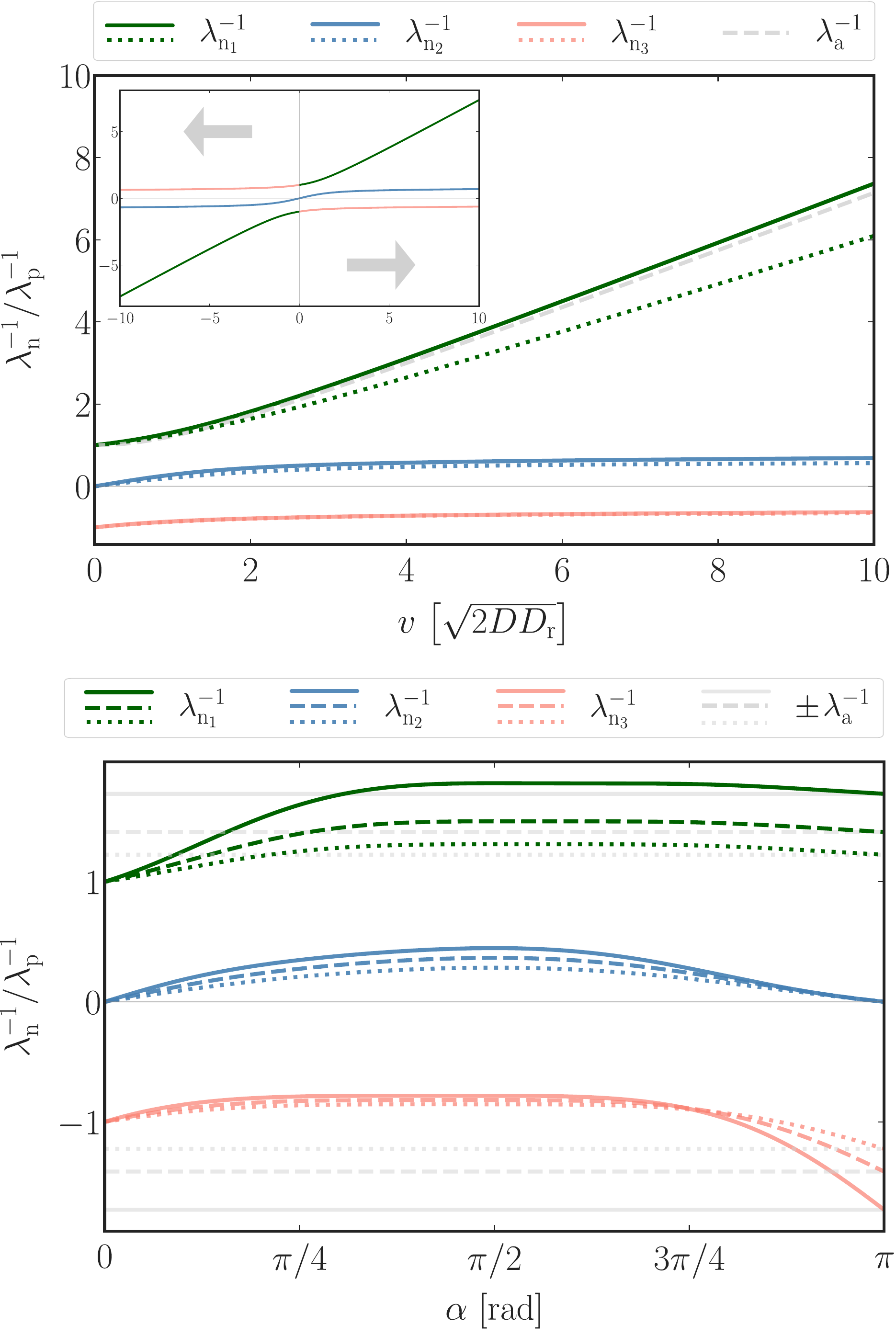}
  \caption{Dependence of eigenvalues $\lambda^{-1}$ on particle propulsion speed $v$ and the nudging acceptance angle $\alpha$, with $\lambda_{\te{n}_i}^{-1}$, $i\in\{1,2,3\}$, denoting the eigenvalues of the matrix $\boldsymbol{\Lambda}$ defined in Eq.~\eqref{eq:X_Lambda_Definition}. The inverse interfacial layer widths $\lambda_{\te{a,p}}^{-1}$ for active/passive layer widths are those from Eq.~\eqref{eq:def_lambda_act-pass}.  \emph{Upper panel:} Eigenvalues were calculated for fixed acceptance angles $\alpha=\ang{45}$ (dotted curves) and $\alpha=\ang{90}$ (solid curves). \emph{Inset:} Dependence of eigenvalues on the nudging direction (indicated by gray arrows). \emph{Lower panel:} Eigenvalues calculated for the fixed P\'{e}clet numbers $\mathcal P \in \{1/2,1,2\}$ (dotted, dashed, solid curve).}
  \label{eigenvalues_lambda}
\end{figure}

We again require that the polarization vanishes for $x \to \pm \infty$, so that the density attains its constant bulk values $\rho_\te{n}=0$  and $\rho_{\te{a,p}}>0$ in the nudging and active/passive bulk, respectively.  Hence, inside the nudging region ($x \leq 0$), only positive eigenvalues $\lambda_{\te{n}_i}^{-1} > 0$ will contribute to the general solution \eqref{eq:general_solution_nudging}.
Since the analytical expressions for the eigenvalues $\lambda_{\te{n}_i}^{-1}$ given in App.~\ref{sec:continuois-model} are not very enlightening, we discuss their behavior graphically in Fig.~\ref{eigenvalues_lambda}.
In both panels, the eigenvalues are measured in units of the inverse characteristic length $\lp^{-1} = \sqrt{\Dr/D}$ of a passive boundary layer.  From the main plot of the upper panel, one infers $\lone \geq \lp^{-1}$, $ \ltwo \geq 0$, $\lthree < 0$,  for all acceptance angles $0 \leq \alpha \leq \pi$ (lower panel and discussion in App.~\ref{sec:comparison}).  Therefore, only $\lone$ and $\ltwo$ contribute to the general solution \eqref{eq:general_solution_nudging}. In contrast to purely active-passive interfaces, which are characterized by a single natural length scale $\lambda_\te{a,p}$ for each side of the activity step, two characteristic lengths $\lambda_1$ and $\lambda_2$ determine the shape of the polarization and density profiles within the nudging layer.
Both $\lone$ and $\ltwo$ grow monotically with increasing propulsion speed $v$; $\ltwo$ less than $\lone$.
While $\ltwo$ remains strictly below the (inverse) natural length $\la^{-1}$ of a fully active polarization layer, $\lone$ might even exceed it, depending on swim speed and acceptance angle.
For slow swim speeds, \ie, small $\mathcal P$, one generally has $\lone > \la^{-1}$, as detailed in App.~\ref{sec:comparison}.

Purely active or passive polarization layers are captured within this framework (eigenvalues of matrix $\boldsymbol{\Lambda}$) as well. In the limiting case of vanishing activity ($v \to 0$), one finds $\ltwo \to 0$, corresponding to a constant (bulk) density in \eqref{eq:general_solution_nudging}, whereas $\lambda_{\te{n}_{1,3}}^{-1} \to \pm \lp^{-1}$.  The positive/negative sign refers to a polarization layer in the negative/positive polarization region. Only one of both eigenvalues contributes whereas the other vanishes for the natural boundary conditions $p(|x|\to \infty) = 0$.  Similarly, for a fully active region ($\alpha \to \pi$), the eigenvalues $\lambda_{\te{n}_{1,3}}^{-1}$ approach $\pm\la^{-1}$, which is indicated by the faint gray lines in the lower panel of Fig.~\ref{eigenvalues_lambda}. Here, positive/negative sign also refers to a polarization layer in the negative/positive region. Again only one of them contributes, due to the boundary conditions. The eigenvalue $\ltwo$ vanishes, corresponding to a constant (bulk) density in Eq.~\eqref{eq:general_solution_nudging}.
The inset of the upper panel of Fig.~\ref{eigenvalues_lambda} contains information about the behavior of all three eigenvalues upon inverting the direction of the nudging process. We find a completely symmetric picture when replacing $v \to -v$.  Only the roles of the eigenvalues change, as we now only allow negative 
eigenvalues to contribute (particles are nudged to the left if $x>0$).  So $\lone$ becomes $\lthree$ and thus does not contribute to $p(x)$ and $\rho(x)$ anymore, while, in return, $\lthree \to \lone$.  The eigenvalue $\ltwo$ changes sign upon inverting the propulsion direction and thus keeps its role.

\begin{figure}[tb!]
  \centering
\includegraphics[width=\columnwidth]{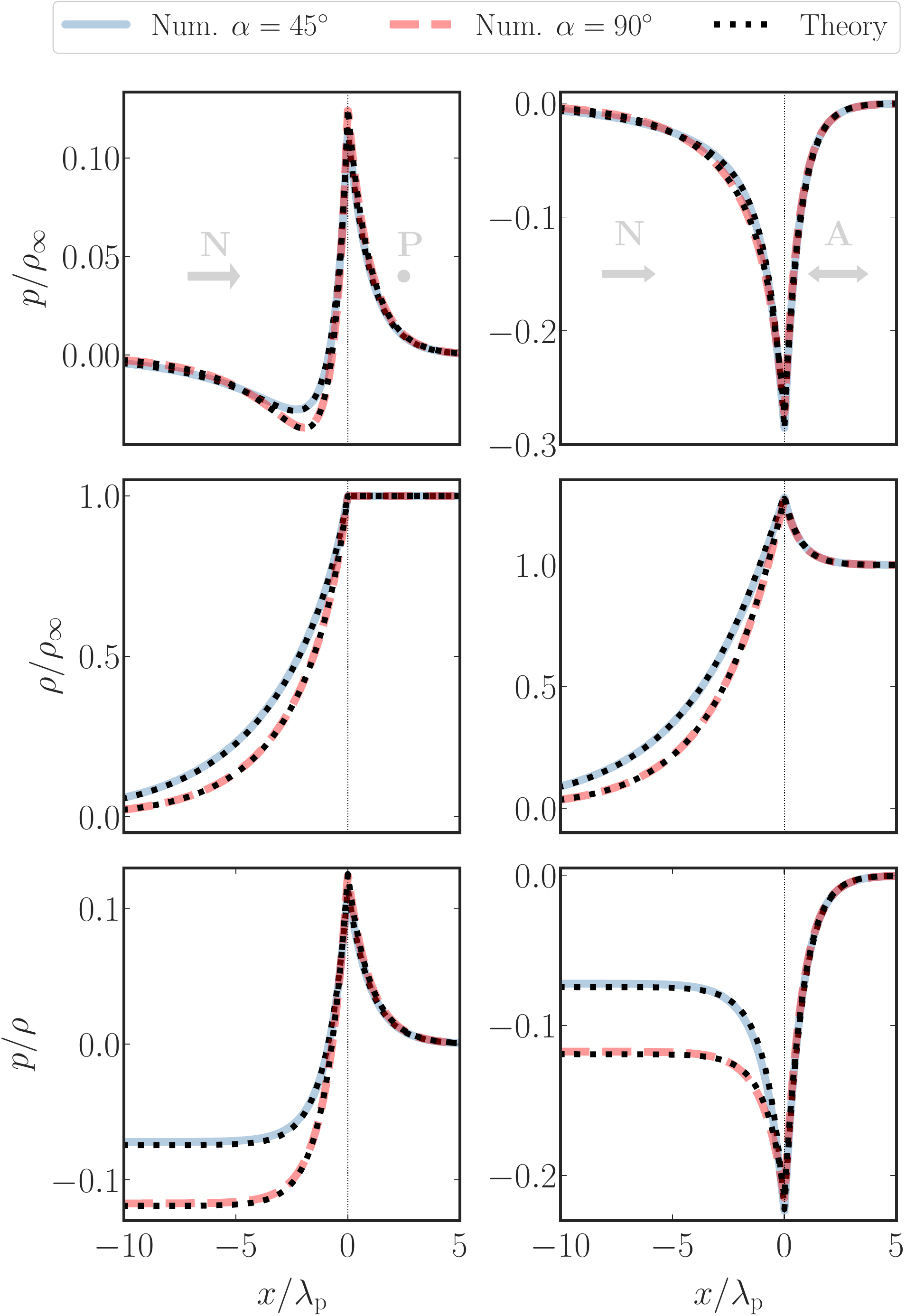}
  \caption{Particle polarization, density, and relative polarization (from top to bottom) according to Eq.~\eqref{eq:general_solution_nudging} versus the exact numerical solution, for $\mathcal P = 1$.  \emph{Left panels:} nudging-passive interface. \emph{Right panels:}  nudging-active interface.   The coefficients $C_i$ are determined by the boundary and matching conditions \eqref{eq:matching_obvious} and \eqref{eq:jump_active_nudging}. }
  \label{fig:pol_rho-nudging}
\end{figure}

Having the eigenvalues $\lambda_{\te{n}_{1,2}}^{-1}$ and the corresponding eigenvectors $\vv w_{1,2}$ of matrix $\boldsymbol{\Lambda}$ that contribute to the general solution \eqref{eq:general_solution_nudging}, we obtained the polarization and density profiles for  nudging-active and nudging-passive interfaces.  Polarization and density profiles are matched to those in the respective passive/active regions by the matching conditions \eqref{eq:matching_obvious} and \eqref{eq:jump_active_nudging}.  The resulting profiles are depicted in Fig.~\ref{fig:pol_rho-nudging} 
for two acceptance angles. 
All our approximate analytical solutions nicely follow the exact numerical results. 
An intuitive physical explanation of the formal density and polarization profiles in terms of a 2-species model is provided in Secs.~\ref{sec:disc-nudg-activ} and \ref{sec:disc-nudg-pass}.

The lower panels of Fig.~\ref{fig:pol_rho-nudging} depict the relative polarization profiles $p(x)/\rho(x)$ in both scenarios. Peaking exactly at the interface, they decay over a length scale (inverse eigenvalue) $\lambda_{\te n_1}<\lambda_{\te n_2}$ into the nudging region (left) and approache a constant nonzero value.  The (relative) bulk polarization $(p/\rho)_{\te n}$ within the nudging region can be calculated explicitly. We therefor rewrite Eq.~\eqref{eq:no_flux_nudging} as
\begin{equation}
  \label{eq:p_by_rho_nudging}
  \frac{p(x)}{\rho(x)}
  =
  \frac{D}{ v(x) \mathcal I_p^{(1)} }
  \frac{\rho'(x)}{\rho(x)}
  -
  \frac{
    \mathcal I_\rho^{(1)}
  }{
    \mathcal I_p^{(1)}
  }.
\end{equation}
Given $\lambda_{\te n_1}<\lambda_{\te n_2}$, and exploiting that the density profile within the nudging region can be written as a linear combination of $\te e^{x/\lambda_{\te n_1}}$ and $\te e^{x/\lambda_{\te n_2}}$ (see Eq.~\eqref{eq:general_solution_nudging} and boundary conditions), one finds $\rho'/\rho \sim \lambda_{\te n_2}^{-1}$ for $|x|$ sufficiently greater than $\lambda_1$. The relative polarization in the nudging bulk is thus given by
\begin{equation}
  \label{eq:p_by_rho_nudging_final}
  \left(
    \frac{p}{\rho}
  \right)_\te{n}
  =
  \frac{
    D
  }{
    v(x) \lambda_{\te n_2} \mathcal I_p^{(1)}
  }
  -
  \frac{
    \mathcal I_\rho^{(1)}
  }{
    \mathcal I_p^{(1)}
  }.
\end{equation}
\begin{figure}[tb!]
  \centering
  \includegraphics[width=\columnwidth]{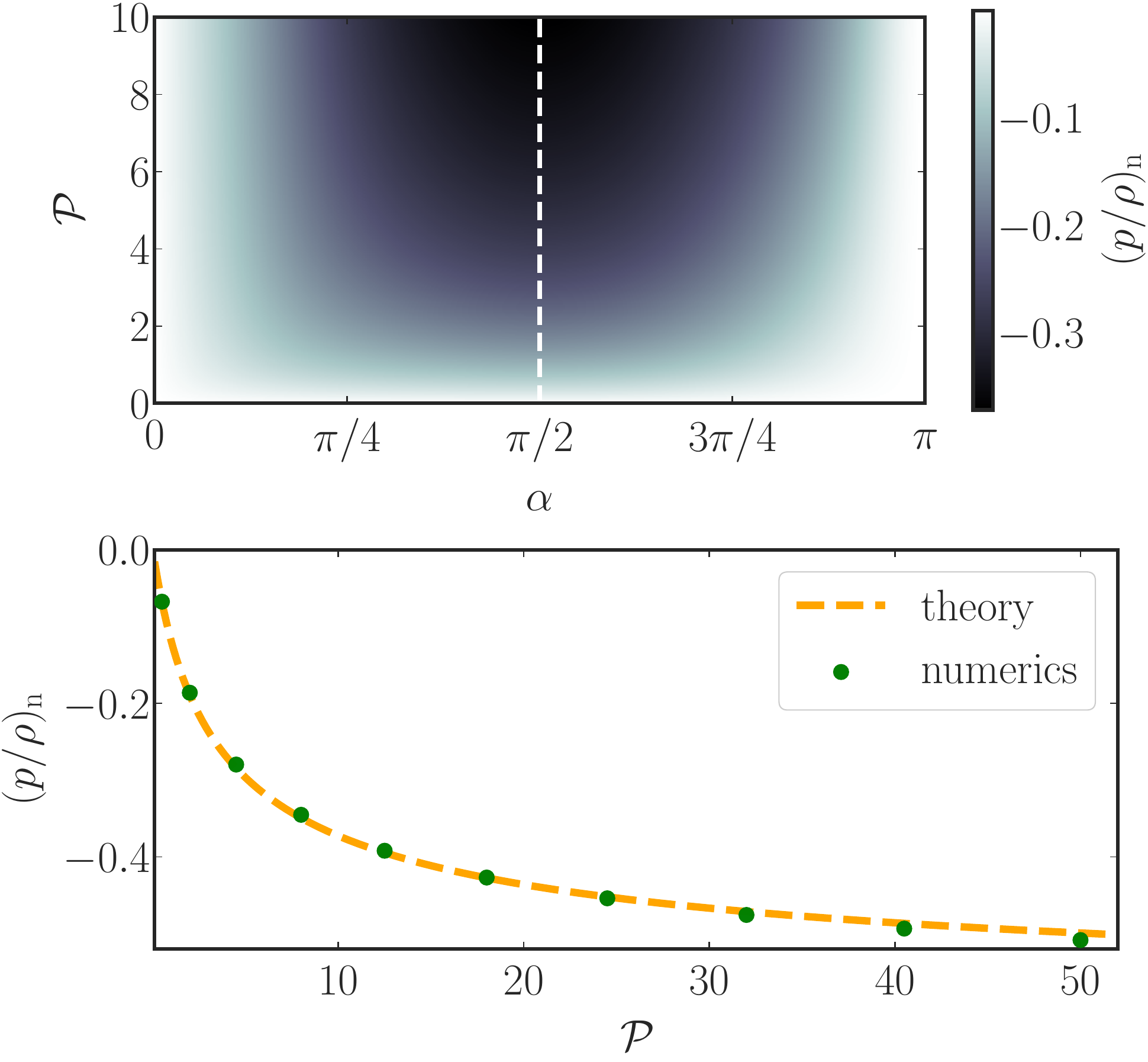}
  \caption{Relative bulk polarization $(p/\rho)_{\te n}$ within the nudging region. \emph{Upper panel:} Heat map of $(p/\rho)_{\te n}$ as function of the P\'{e}clet number $\mathcal P$ and acceptance angle $\alpha$ according to Eq.~\eqref{eq:p_by_rho_nudging_final}. The vertical dashed line depicts the extremum (minimum) of $(p/\rho)_{\te n}$ with respect to $\alpha$. \emph{Lower panel:} approximate theory  \eqref{eq:p_by_rho_nudging_final} (dashed line) versus exact numerical solution (circles) for  $\alpha=\pi/2$.}
  \label{fig:bulk_polarization_nudging}
\end{figure}
For a nudging layer in the region with $x<0$, one finds that $(p/\rho)_{\te n}\leq 0$ and that it becomes extremal for $\alpha = \pi/2$, as can be inferred from the upper panel of Fig.~\ref{fig:bulk_polarization_nudging}. Also, the higher the swimmer's activity, the higher the extremum in the bulk polarization. The lower panel of Fig.~\ref{fig:bulk_polarization_nudging} displays a convincing agreement between the approximate theoretical prediction \eqref{eq:p_by_rho_nudging_final} and exact numerical solutions up to P\'{e}clet numbers $\mathcal P$ on the order of 50. We attribute the subsequent growing mismatch between both to the fact that there exists no one-to-one mapping between the (approximate) analytical theory and the simple 2-state model in the case of a nudging interface (see Sec.~\ref{sec:nudging}).

As explicitly checked above and proven to hold generally \cite{hermann2020PolStateFct}, the total polarization
\(
P_\te{tot}
=
\int_{-\infty}^\infty \df x ~ p(x)
\)
is determined by the difference
of the respective bulk fluxes $v_i \rho_i$ divided by $2 \Dr$; cf.~Eq.~\eqref{eq:Pol_total}.
The bulk fluxes associated with the nudging and the passive region are both zero. The former vanishes because of $\rho_\te{n} = 0$ in the bulk, the latter because of $v = 0$.   
 Thus, for a nudging-passive interface, $P_\te{tot} = 0$.  Vanishing bulk fluxes therefore explain the change of sign in the polarization profile for the nudging-passive interface (see upper left panel of Fig.~\ref{fig:pol_rho-nudging}). For the nudging-active interface, one finds $P_\te{tot} = - \rho_\te{a} \va /(2 \Dr)$, which coincides with the total polarization at a passive-active interface; cf.~Eq.~\eqref{eq:Pol_total}. We verified these results for $P_\te{tot}$ at the considered nudging interfaces analytically and numerically.

 Note that merely the total polarization is determined by bulk quantities, but not the local polarization profile $p(x)$ itself. A thorough discussion and physical interpretation of the (active, passive, nudging) polarization layers is provided in the following section.

\section{Intuitive physical interpretation}
\label{sec:disc-bound-layer}

\subsection{The 2-species run-and-tumble model}
\label{sec:2-species-model}

The emerging polarization and inhomogeneous density distribution in the vicinity of an activity step are easily understood within a simple one-dimensional 2-species run-and-tumble model \cite{Cates2012DiffusivePhysics,Tailleur2008StatisticalBacteria,Malakar2018RunAndTumble1D, Weiss_2002,razin2020entropyProdABPinBox}. The swimmer is obliged to either orient parallel or anti-parallel to the $x$-axis but can randomly flip its orientation (``tumble'') at a rate $k$  (corresponding to a dichotomous Markov process). During the ``run phases'', besides its thermal diffusion, the particle propels actively with a position dependent swim speed $v_\pm(x)$, which might also depend on the orientation $(\pm)$ of the particle in order to mimic the nudging. We define the probability densities $n_+(x,t)$ and $n_-(x,t)$ for encountering the particle at time $t$ at position $x$ with orientation parallel ($+$) or anti-parallel ($-$) to the $x$-axis.
The corresponding fluxes $J_\pm(x,t)$ contain contributions both from thermal agitation and from active propulsion, and are given by
\begin{align}
  \label{eq:J+}
  J_+(x,t)
  &=
    -D n_+'(x,t) + v_+(x) n_+(x,t),
  \\[0.5em]
  \label{eq:J-}
  J_-(x,t)
  &=
    -D n_-'(x,t) +v_-(x) n_-(x,t).
\end{align}
For the (total) density $\rho \equiv n_+ + n_-$ and the polarization $p \equiv n_+- n_-$ we keep the notation of the continuous-angle model.
In the steady state, the total flux $J_+ + J_-$ vanishes, which yields the balance condition
\begin{equation}
  \label{eq:flux_balance_2spec}
  D \rho'
  =
  \frac{v_++v_-}{2}\rho
  +
  \frac{v_+-v_-}{2}p.
\end{equation}
The time evolution of the densities is given by
\begin{align}
  \dot n_+
  &=
    \nonumber
    -J_+'
    -
    k (n_+ - n_-)
  \\
  &=
    \label{eq:n+_dot}
    D n_+''
    -
    (v_+ n_+)'
    -
    k (n_+ - n_-),
  \\[0.5em]
  \dot n_-
  &=
    \nonumber
    -J_-'
    +
    k (n_+ - n_-)
  \\
  &=
    \label{eq:n-_dot}
    D n_-''
    -
    (v_- n_-)'
    +
    k (n_+ - n_-).
\end{align}
In the steady state $(\dot n_\pm =0)$, subtracting Eq.~\eqref{eq:n-_dot} from Eq.~\eqref{eq:n+_dot}  yields
\begin{equation}
  \label{eq:ode_pol_2spec}
  p''
  =
  \frac{2k}{D}p
  +
  \left(
    \frac{v_+-v_-}{2D} \rho
    +
    \frac{v_++v_-}{2D} p
  \right)'.
\end{equation}

\begin{figure}[tb!]
  \centering
  \includegraphics[width=\columnwidth]{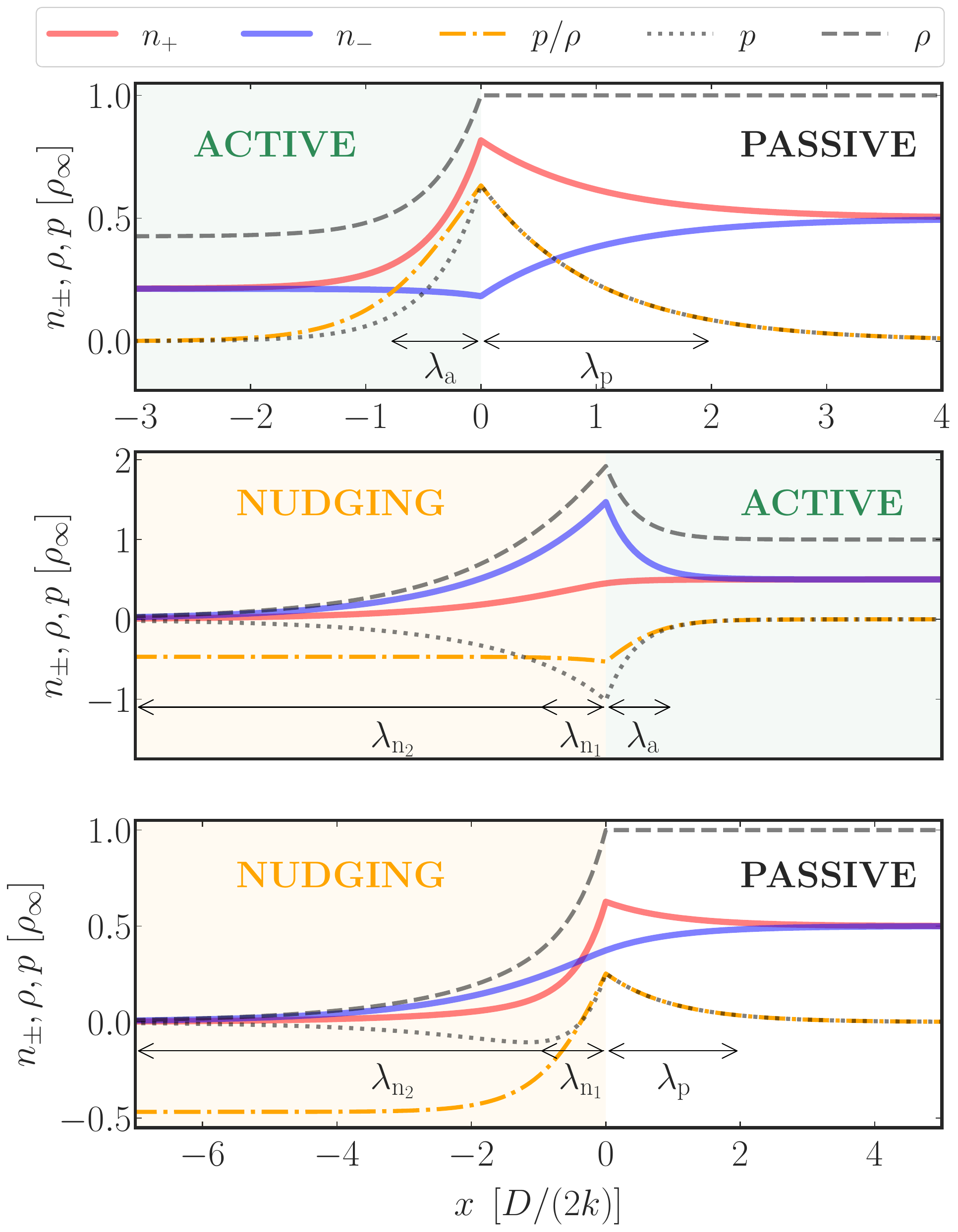}
  \caption{Exact analytical solution of the 2-species model: polarization and density profiles, $p(x)$ and $\rho(x)$, their ratio $p/\rho$, and the species concentrations $n_\pm(x) = [\rho(x) \pm p(x)]/2$ of right- and left-oriented particles around an active-passive, nudging-active, and nudging-passive interface located at $x=0$. All quantities are normalized by the respective constant bulk density $\rho_\infty = \rho(x\to\infty)$. The natural widths of the interfacial polarization layers  are denoted by $\la$, $\lp$, and $\lambda_{\te{n}_{1,2}}$ in the active, passive, and nudging regions, respectively.}
  \label{fig:2spec_model}
\end{figure}

\paragraph*{Fully active/passive:}
\label{sec:fully-activepassive:}
The case of a symmetrically active/passive particle (regardless the orientation) is captured by setting $v_+ \equiv v = -v_-$. Equations \eqref{eq:flux_balance_2spec} and \eqref{eq:ode_pol_2spec} then reduce to
\begin{equation}
  \label{eq:no_flux_2spec_symm}
  \rho'
  =
  \frac{v}{D} p,
  \qquad
  p''
  =
  \frac{p}{\lambda^2}
  +
  \frac{\rho}{D}v',
\end{equation}
with the characteristic length scale
\begin{equation}
  \label{eq:lambda_2spec}
  \lambda(x)
  \equiv
  \left[
    \frac{2k}{D}
    +
    \frac{v^2(x)}{D^2}
  \right]^{-1/2}.
\end{equation}
The above equations for $\rho$, $p$, and $\lambda$ are structurally equal to Eqs.~\eqref{eq:flux_balance_act-pass}, \eqref{eq:ode_pol_act-pass} and \eqref{eq:def_lambda_act-pass} in the (approximate) model for an active particle that can rotate continuously in the plane.  Upon  mapping $2k \to \Dr$, $v \to v/\sqrt{2}$, $\rho \to \rho/\sqrt{2}$, the two models become equivalent. One therefore applies the same methods as above to obtain the analytical solutions for $p(x)$ and $\rho(x)$. Both functions as well as the species densities $n_\pm(x) = [\rho(x) \pm p(x)]/2$ and the relative polarization $p/\rho$ are plotted in the upper panel of Fig.~\ref{fig:2spec_model}. All quantities are normalized by the bulk density $\rho_\infty = \rho(x\to\infty)$ in the passive region.

\paragraph*{Nudging:}
\label{sec:nudging}

Within the framework of the 2-species model, a nudging process is modelled by setting $v_+ \equiv v$ and $v_- \equiv 0$ for nudging to the right or \emph{vice versa} for nudging to the left. The flux-balance condition \eqref{eq:flux_balance_2spec} and Eq.~\eqref{eq:ode_pol_2spec} for the polarization then become
\begin{align}
  \label{eq:flux_balance_2spec_nudging}
  \rho'
  &=
  \frac{v}{D} \frac{\rho}{2}
  +
    \frac{v}{D} \frac{p}{2},
  \\[0.5em]
  \label{eq:ode_pol_2spec_nudging}
  p''
  &=
    \frac{2k}{D}p
    +
    \frac{v}{D}
    \left(
    \frac{\rho'}{2} + \frac{p'}{2}
    \right)
    +
    \frac{v'}{D}
    \left(
    \frac{\rho}{2} + \frac{p}{2}
    \right).
\end{align}
These equations also have the same structure as their counterparts \eqref{eq:no_flux_nudging} and \eqref{eq:ode_pol_nudging} for the continuous-rotation model. Thus, the methods used in section \ref{sec:photon-nudging} can be applied in order to determine the polarization and density profiles.  Note, however, that here, the 2-species and the continuous-angle model can not be mapped onto each other. There exists no unique acceptance angle $\alpha$ to ensure
\(
v/2 = \va \mathcal I_{(\rho,p)}^{(1,2)}
\)
for all coefficients $\mathcal I_{(\rho,p)}^{(1,2)}$ defined in \eqref{eq:I_rho_1}-\eqref{eq:I_p_2}, where $\va$ denotes the Janus swimmer's propulsion speed in the continuous-angle model.  Nevertheless, at least qualitatively, polarization and density profiles in the 2-species model display the same features as their counterparts in the continuous-angle model, as can be inferred from the second and third panel of Fig.~\ref{fig:2spec_model} (cf.~Fig.~\ref{fig:pol_rho-nudging}). A more thorough discussion of the nudging process within the framework of the 2-state model, and a comparison to the continuous model can be found in App.~\ref{app:2-species-model} and \ref{sec:comparison}.

\subsection{Active-passive interface}
\label{sec:discussion}

To gain an intuitive understanding of the emerging polarization layer at an active-passive interface, focus on the upper panel of Fig.~\ref{fig:2spec_model}. For the sake of brevity we denote particles pointing to the left and right by $L$ and $R$, respectively. We first consider the situation on the active side, where the particle motion might be regarded as quasi-ballistic. While $R$-particles get stuck at the  interface due to the ceasing propulsion, $L$-particles quickly ``escape'' the interfacial region. This sorting mechanism leads to a majority of $R$- particles at the interface caused by hidden bulk currents \cite{hermann2020PolStateFct}. We now qualitatively explain the shape two polarization layers. In the passive region, the characteristic decay length $\lp = \sqrt{D/(2k)}$ of the polarization is obtained by setting $v(x) = 0$ in Eq.~\eqref{eq:lambda_2spec}. It is intuitively understood since the particle's motion on the passive side is ordinary diffusion with diffusivity $D$. The spreading of the excess polarization into the passive region is limited by the  characteristic flipping time $(2k)^{-1}$.
Therefore, $\lp$ basically coincides with the mean-squared displacement of a passive particle during this time.  On the active side, a kind of ``sedimentation pressure'' joins the game. Its cause is an ``active swim force'' $\zeta v$ (with friction $\zeta$) directed towards the interface for the $R$-particles and away from it for $L$-particles. One thus might regard  $R$-particles as ``heavy'' and $L$-particles as ``buoyant''. The effective sedimentation process therefore compresses the extension $\la$ of the polarization layer according to Eq.~\eqref{eq:lambda_2spec} on the active side, which, for highly persistent motion ($v^2/(2kD) \gg 1$), is described by the barometer formula $\exp(-v \zeta x/ \zeta D)$. The swim force takes the role of gravity and $\zeta D$ of thermal energy $\kB T$, according to the Sutherland--Einstein relation. In the same limit follows the motility-induced density suppression $\Ra/\Rp = \lp/\la \propto v^{-1}$, which is a well-known result for run-and-tumble particles \cite{Schnitzer1993TheoryChemotaxis,Cates2012DiffusivePhysics}.

\subsection{Nudging-active interface}
\label{sec:disc-nudg-activ}

The middle panel of Fig.~\ref{fig:2spec_model} depicts $p(x)$, $\rho(x)$, their ratio, and the species concentrations $n_\pm(x)$ in the vicinity of a nudging-active interface. That is, while both particle species perform persistent motion within the active region, only the $R$-particles propel actively inside the nudging region. The $R$-species therefore does not display a sudden change (kink) in its species concentration $n_+(x)$ upon crossing the interface, as no abrupt activity drop is experienced.  In the active region, $R$-particles quickly ``escape'' the interfacial area into the active region whereas $L$-particles get stuck at the interface and venture only diffusively into the nudging region. We thus observe an excess of $L$-particles at the interface, and therefore a negative polarization. The extent $\la$ of the polarization layer on the active side is again determined by the interplay between Brownian motion and the effective sedimentation pressure, this time with ``heavy'' $L$- and ``buoyant'' $R$-particles.  The symmetry between ``heavy'' and ``buoyant'' particle species is broken in the nudging region. There, $R$-particles are ``heavy'' and therefore nudged towards the interface  whereas the $L$-species are passive Brownian particles.
Close to the interface, the decay of density and polarization into the nudging region is therefore characterized by a length scale $\lambda_{\te{n}_1} \neq \la$. Referring back to the upper panel of Fig.~\ref{eigenvalues_lambda}, we infer that $\lambda_{\te{n}_1}$ is still quite similar to the characteristic decay length $\la$ pertaining to a purely active region.
Notice that, over a distance  $\lambda_{\te{n}_1}$, the relative polarization $p/\rho$ approaches a constant value given by Eq.~\eqref{eq:p_by_rho_nudging_final}. Which is a distinctive feature of the bulk in the nudging region, characterized by the swim speed and the acceptance angle of the nudging procedure. In contrast to purely active or passive regions, both the particle polarization $p(x\to\infty)$ and bulk density $\rho(x\to\infty)$ in the nudging region decay to zero, since every particle is inevitably nudged towards the interface until it crosses it. The decay of the absolute polarization and density profiles towards zero is described by the second characteristic length scale $\lambda_{\te{n}_2} > \lambda_{\te{n}_1}$.

\subsection{Nudging-passive interface}
\label{sec:disc-nudg-pass}

Finally, consider the polarization and density profiles and the respective species concentrations presented in the last panel of Fig.~\ref{fig:2spec_model}. Now, only $R$-particles propel actively inside the nudging region, while both particle species behave like ordinary Brownian particles in the passive region. Therefore, the $L$-particles can cross the interface smoothly and their density $n_-(x)$ does not display a kink.
We observe an excess of $R$-particles at the interface by virtue of the inherently biased nudging process, and thus a positive polarization. The width $\lp$ of the polarization layer on the passive side is determined by the distance $\sqrt{D/(2k)}$ covered by thermal diffusion during the characteristic time scale $(2k)^{-1}$ for ``tumbling''. The imbalance between the ``heavy'' R- and neutral $L$-particles determines the spreading of the polarization into the nudging region. Due to the ``removal'' of $R$-particles towards the interface by virtue of the nudging procedure, the polarization even changes sign and becomes negative over the characteristic length scale $\lambda_{\te{n}_1}$ before it converges to zero over the length scale $\lambda_{\te{n}_2}$ as discussed in the previous scenario. This distinctive shape of the polarization can be seen as indicative of  the hidden bulk currents that are generally understood to cause the interfacial polarization layers \cite{hermann2020PolStateFct}, as already discussed in Sec.~\ref{sec:photon-nudging}.

\section{Conclusion}
\label{sec:conclusion}
In this article, we have studied the behavior of a single Janus-type swimmer in the vicinity of a motility step. Within an approximate ABP model, we derived analytical expressions for the polarization and density profiles of a Janus particle at planar activity steps. We showed that they agree well with exact numerical solutions and experimental data (see our companion article \cite{Soeker2020ActivityFieldsPRL}). Key features of polarization and density profiles at motility steps were discussed and shown to exhibit important similarities to those observed for MIPS. As a consistency check, we also explicitly demonstrated both analytically and numerically that the total polarization induced by the motility step is determined by the difference of (hidden) bulk fluxes, and thus constitutes a state function. We further showed that the bulk density ratio between two regions of distinct but constant activity is determined by the ratio of the respective effective diffusion coefficients and independent of the shape of the activity profile that mediates between the bulk regions. Motivated by the versatile experimental technique of photon nudging, we moreover generalized our theoretical results to the situation of orientation-dependent propulsion speeds.
We conclude that the co-localization of polarization and density patterns in activity gradients, as they naturally occur at various interfaces, is a characteristic phenomenological trait to robustly distinguish motile-particle suspensions from thermal and athermal passive suspensions.

  \section*{Acknowledgements}
\label{sec:acknowledgements}

We thank Paul Cervenak and Anton Stall for discussions and  contributions during the early stages of this work.
We acknowledge funding by the Deutsche Forschungsgemeinschaft (DFG) via SPP 1726/1 and KR 3381/6-1, and by Czech Science Foundation (project No. 20-02955J).
Viktor Holubec gratefully acknowledges support by the Humboldt foundation.

\bibliography{references}

\appendix

\section{Planar activity step}
\label{sec:planar-activity-step}

\subsection{Density ratio and total polarization}
\label{sec:dens-ratio-total}

\paragraph{Density ratio:}
Introducing the auxiliary quantities
\(
\beta_{\te{A},\te{a}}
\equiv
p_\te{max} v_{\te{A},\te{a}} \lambda_{\te{A},\te{a}} /[D \rho(0)],
\)
Eqs.~\eqref{eq:rho_step<} and \eqref{eq:rho_step>} evaluated at $x=0$ and $x\to\infty$, respectively, yield 
\begin{align}
  \label{eq:app_rho_step<}
  \rho(0) - \rho_\te{A}
  &=
    \rho(0) \beta_\te{A}
  \\[0.5em]
  \label{eq:app_rho_step>}  
  \rho_\te{a} - \rho_\te{A}
  &=
    \rho(0) (\beta_\te{a} + \beta_\te{A}).
\end{align}
Using these equations, the density ratio can be expressed as
\(
\rho_\te{a}/\rho_\te{A}
=
(1+\beta_\te{a}) / (1-\beta_\te{A})
\).  
Further using
\begin{align}
  \label{eq:app_lambda}
  \lambda_{\te{A},\te{a}}
  &=
    \left(
    \frac{\Dr}{D} + \frac{v_{\te{A},\te{a}}^2}{2D^2}
    \right)^{-1/2},
  \\[0.5em]
  \label{eq:app_pmaxbyrho0}
  \frac{p_\text{max}}{\rho(0)}
  &=
  \frac{\vA - \va}{2D}
  \frac{\lA \la
  }{
    \lA + \la
  },
\end{align}
from Eq.~\eqref{eq:pol_peak_nat-bound}, implies
\(
\rho_\te{a} / \rho_\te{A}
=
\la / \lA,
\)
as given in Eq.~\eqref{eq:dentiy_ratio_nat-bound}.

\paragraph{Total Polarization:}

From Eq.~\eqref{eq:Pol_total}
we know that the total polarization $P_\te{tot}$ is given by
\(
p_\te{max} (\lA + \la).
\)
The coefficient $p_\te{max}$ can be expressed as
\(
D(\rho_\te{a} - \rho_\te{A}) / (\vA \lA - \va \la)
\)
by virtue of Eq.~\eqref{eq:app_rho_step>}. Hence, the total polarization reads
\begin{equation}
  \label{eq:app_Ptot_1}
  P_\te{tot}
  =
  D (\rho_\te{a} - \rho_\te{A})
  \frac{
    \la + \lA
  }{
    \vA \lA + \va \la
  }.
\end{equation}
Using
\(
\rho_\te{a} / \rho_\te{A}
=
\la / \lA,
\)
we find that
\begin{align}
  P_\te{tot}
  &=
    D
    \frac{
    (\rho_\te{a} - \rho_\te{A}) (\rho_\te{a} + \rho_\te{A})
    }{
    \vA \rho_\te{A} + \va \rho_\te{a}
    }
  \\[0.5em]
  &=
    D
    \frac{
    \rho_\te{a}^2 - \rho_\te{A}^2
    }{
    \va^2 \rho_\te{a}^2 - \vA^2 \rho_\te{A}^2
    }
    (\va \rho_\te{a} - \vA \rho_\te{A}).
\end{align}
Using the formula \(
\rho_\te{a} / \rho_\te{A}
=
\la / \lA
\)
and the definition \eqref{eq:app_lambda} of $\lambda_{\te{A},\te{a}}$, the factor in front of the term in parentheses turns out to be equal to $-1/(2\Dr)$. One thus obtains
\(
P_\te{tot}
=
(\vA \rho_\te{A} - \va \rho_\te{a})/(2\Dr),
\)
as stated in Eq.~\eqref{eq:Pol_total}.

\subsection{Vanishing integral}
\label{sec:vanishing-integral}

The main result \eqref{eq:formal_solution_density_ratio} for the density ratio reads
\begin{align}
  \label{eq:density_ratio_app}
  \frac{\rho(x)}{\rho(x_0)}
  &=
  \sqrt{ \frac{\Deff(x_0)}{\Deff(x)} } ~
  \te{exp}
    \left\{
    \mathcal U[v](x_0,x)
    \right\},
\end{align}
with the emerging functional $\mathcal U$ defined in Eq.\eqref{eq:density_ratio_functional}.
We will first prove that for the ratio of bulk densities, $\mathcal U \equiv 0$, within the framework of the 2-state model (see Sec.~\ref{sec:2-species-model}). After that, we will describe how this proof generalizes to the continuous model.

\begin{figure}[tb!]
  \centering
  \includegraphics[width=\linewidth]{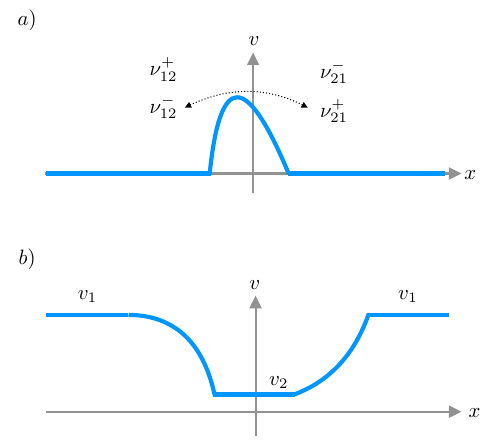}
  \caption{Activity profiles $v(x)$ considered to demonstrate that the integral $\mathcal U$ vanishes. In the upper plot, $\nu_{ij}^\pm$ denote the transition rates of particles species $(\pm)$ from the left to the right bulk region and vice versa.  In the lower plot, $v_{1,2}$ denote the constant activities in the middle and the outer regions, respectively.}
  \label{fig:proof_integral}
\end{figure}

First, we consider an arbitrary activity profile $v(x)$ whose inhomogenities are localized around a finite region beyond which the velocity assumes a single constant value, see Fig.~\ref{fig:proof_integral} a).  We will refer to the corresponding constant bulk densities as $\rho_i$, $i=1,2$. In the bulks, the polarization vanishes and thus the concentrations $n_\pm^{(i)}$ of the individual species equal, $n_\pm^{(i)} = \rho_i/2$. The corresponding fluxes between the two bulk regions can be expressed as
\begin{align}
  \label{eq:J12}
  J_{12}
  &=
  \frac{\rho_1}{2}
  \left(
  \nu_{12}^+ + \nu_{12}^-
  \right),
  \\[0.5em]
  \label{eq:J21}
  J_{21}
  &=
  \frac{\rho_2}{2}
  \left(
  \nu_{21}^+ + \nu_{21}^-
  \right),
\end{align}
where we introduced the transition rates $\nu_{ij}^\pm$ of particle species $(\pm)$ between the bulk region $i$ and $j$.  In the steady state, the (anti-)symmetry between the two particle species implies $\nu_{12}^+ = \nu_{21}^-$ and $\nu_{12}^-=\nu_{21}^+$.  Flux balance $J_{12} = J_{21}$ then induces $\rho_1 = \rho_2$.  
For an arbitrary activity profile $v(x)$ with bulk activities given by a single constant, Eq.~\eqref{eq:density_ratio_app} thus implies that $1 = \rho_1 /\rho_2 = \rm e^\mathcal U$ and hence $\mathcal U = 0$.

Now, we consider activity profiles of the form sketched in Fig.~\ref{fig:proof_integral} b). Two bulk region with constant activity $v_1$ are interconnected by two arbitrary activity steps and an intermediate bulk region with a constant activity $v_2\neq v_1$.
From the first part of the proof, we know that $\rho_1/\rho_3 = 1$ as the activity in both regions is equal and constant. We thus know that $0 = \mathcal U[v_1,v_2] = \mathcal U_1[v_1] + \mathcal U_2[v_2]$, where we introduced the integrals $\mathcal U_{1,2}$ pertaining to the two activity profiles $v_{1,2}$.  Since the total integral has to vanish for any intermediate set of activity gradients, both $\mathcal U_1$ and $\mathcal U_2$ must be zero.
Therefore, for the 2-state model,
\(
\rho_1 / \rho_2
=
\sqrt{\Deff(x_2)/\Deff(x_1)}
\)
for any activity profile that mediates between two bulk positions $x_1$ and $x_2$.

To generalize this proof to the model with continuous orientation, it is enough to realize that for each orientation pointing to the right there exists an ``anti'' particle with orientation pointing to the left. These two particle orientation can be considered as representatives of our two species, and, to finish the proof, it is thus enough to apply the above procedure to the whole (infinite) set of twin orientations.


 \subsection{Finite system -- determining the coefficients}
 \label{sec:integration_constants}
 
The general solutions \eqref{eq:pol_act_pass_general_finite} and \eqref{eq:rho_solution_active_finite} to the particle's density distribution $\rho_{\rm{a/p}}(x)$ and polarization $p_{\rm{a/p}}(x)$ within the active ($0 \leq x \leq \xif$) and passive ($a < x \leq L$) region, respectively, read
\begin{align}
  \label{eq:app_polarization_active}
  \Pa(x)
  &=
    \Ca
    \sinh
    \left(
    \frac{x}{\la}
    \right),
    \quad
    \la
    =
    \left(
    \frac{\Dr}{D}
    +
    \frac{\va^2}{2D^2}
    \right)^{-1/2}
  \\[0.5em]
  \label{eq:app_polarization_passive}
  \Pp(x)
  &=
    \Cp
    \sinh
    \left(
    \frac{L-x}{\lp}
    \right),
    \quad
    \lp
    =
    \sqrt{\frac{D}{\Dr}},
  \\[0.5em]
  \label{eq:app_density_active}
  \rho_{\rm a}(x)
  &=
    \rho_{\rm a}(0)
    +
    \frac{\Ca \la \va}{D}
    \left[
    \cosh
    \left(
    \frac{x}{\la}
    \right)
    -
    1
    \right]
  \\[0.5em]
  \label{eq:app_density_passive}
  \rho_{\rm p}(x)
  &=
  \rho_{\rm a}(\xif)
  = const.
\end{align}
The emerging integration constants $\Ca$ and $\Cp$ are fixed by the matching conditions
\begin{align}
  \Pp(\xif)
  =
  \Pa(\xif),
  \\[0.5em]
  \label{eq:app_jump_cond_act-pass}  
  \Pa'(\xif) - \Pp'(\xif)
  =
  \frac{\va}{2D} \rho(\xif),
\end{align}
which follow from Eq.~\eqref{eq:cont_condition_general}.
The first one allows us to express one integration constant in terms of the other:
\begin{equation}
  \label{eq:ca_in_terms_of_cp}
  \Cp
  =
  \Ca
  \frac{
    \sinh[\xif/\la]
  }{
    \sinh[(L-\xif)/\lp]
  }.
\end{equation}
Before we employ the condition \eqref{eq:app_jump_cond_act-pass}, we exploit the normalization condition
\(
\int_{-L}^L \df x ~ \rho(x) = 1
\)
to calculate the density at the interface $\rho(\xif)$.
Respecting the symmetry of the considered problem, we get
\begin{widetext}
  \begin{align}
    \nonumber
    \frac12
    =
    \int\limits_0^L \df x ~\rho(x)
    &=
      \int\limits_0^{\xif} \df x ~\rho_{\rm a}(x)
      +
      \int\limits_\xif^L \df x ~\rho_{\rm p}(x)
    \\[0.5em]
    \nonumber
    &=
      \int\limits_0^{\xif} \df x ~\rho_{\rm a}(x)
      +    
      (L-\xif) \rho_{\rm a}(\xif)
    \\[0.5em]
    \label{eq:app_rho_norm}
    &=
      \xif \rho_{\rm a}(0)
      +
      \frac{\Ca \la \va}{D}
      \left[
      \la \sinh
      \left(
      \frac{\xif}{\la}
      \right)
      -
      \xif
      \right]
      +
      (L-\xif)
      \left\{
      \rho_{\rm a}(0)
      +
      \frac{\Ca \la \va}{D}      
      \left[
      \cosh
      \left(
      \frac{\xif}{\la}
      \right)
      -
      1
      \right]
      \right\}.
  \end{align}      
Solving this equation for $\Ra(0)$, we find
\begin{equation}
  \rho_{\rm a}(0)
  =
  \frac{1}{2L}
  -
  \left(
    1 - \frac{\xif}{L}
  \right)
  \frac{\Ca \la \va}{D}
  \left[
    \cosh
    \left(
      \frac{\xif}{\la}
    \right)
    -
    1
  \right]
  -
  \frac{\xif}{L}
  \frac{\Ca \la \va}{D}
  \left[
    \frac{\la}{\xif}
    \sinh
    \left(
      \frac{\xif}{\la}
    \right)
    -
    1
    \right].
\end{equation}
Substituting this relation into Eq.~\eqref{eq:app_density_active} yields
\begin{equation}
  \rho_{\rm a}(\xif)
  =
  \frac{1}{2L}
  +
  \frac{\xif}{L}
  \frac{\Ca \la \va}{D}
  \left[
  \cosh
  \left(
  \frac{\xif}{\la}
  \right)
  -
  \frac{\la}{\xif}
  \sinh
  \left(
  \frac{\xif}{\la}
  \right)
  \right].
\end{equation}
Finally, plugging this result into Eq.~\eqref{eq:app_jump_cond_act-pass} and using Eq.~\eqref{eq:ca_in_terms_of_cp} for $\Cp$ renders a linear equation for $\Ca$, which is straightforwardly solved. With the definitions \eqref{eq:Pmax}-\eqref{eq:r_rho},
\begin{align}
  \label{eq:Pmax_rho}
  P_{\te{max}}
  \equiv
  \frac{\va}{2D}
  \frac{\la \lp}{\la + \lp},
  \qquad
  r_\rho
  \equiv
  \frac{\la}{\lp},
\end{align}
 one finds that
\begin{equation}
  \label{eq:Ca_general}
  \Ca
  =
  \frac{1}{2 L \sinh(\xif/\la)}
  \frac{P_{\te{max}}
  }{
    \coth
    \left(
      \frac{L-\xif}{\lp}
    \right)
    -
    \frac{1-r_\rho}{L}
    \left[
      \xif \coth
      \left(
        \frac{L-\xif}{\lp}
      \right)
      -
      \la
    \right]
  }.
\end{equation}
The approximation
\(
\coth((L-\xif)/\lp)
\approx
1
\)
in the above equations then leads to the polarization and density profiles \eqref{eq:pol_compact_act}-\eqref{eq:rho_compact_pass} given in the main text.
\end{widetext}

\section{Nudging Layer}
\label{sec:nudging-layer}

\subsection{Continuous-angle model}
\label{sec:continuois-model}

As derived in Sec.~\ref{sec:photon-nudging}, within a nudging region, the vector
\(
  \vv X(x)
  =
  \left[
    p'(x),p(x),\rho(x)
  \right]^\top
  \)
  composed of the polarization profile $p(x)$, its derivative, and the density $\rho(x)$
  obeys an equation of the form
  \(
  \vv X'
  =
  \boldsymbol{\Lambda} \vv X.
  \)
  In contrast to the discussion in the main text we now resort to a dimensionless description by introducing $\lp=\sqrt{D/\Dr}$ as natural unit of length.
The matrix $\boldsymbol{\Lambda}$ can then be expressed in terms of a single parameter -- the P\'{e}clet number $\mathcal P = \va^2/(2 D \Dr)$ -- and reads
\begin{equation}
  \label{eq:Lambda_Definition}
  \boldsymbol{\Lambda}
  =
  \begin{pmatrix}
    \sqrt{2 \mathcal P} \mathcal I_p^{(2)}
    &
    1
    +
    2 \mathcal P
    \mathcal I_\rho^{(2)} \mathcal I_p^{(1)}
    &
    2 \mathcal P
    \mathcal I_\rho^{(1)} \mathcal I_\rho^{(2)}
    \\[0.2em]
    1 & 0 & 0
    \\[0.2em]
    0
    &
    \sqrt{2 \mathcal P} \mathcal I_p^{(1)}
    &
    \sqrt{2 \mathcal P} \mathcal I_\rho^{(1)}
  \end{pmatrix}
\end{equation}
The quantities $\mathcal I_{\rho,p}^{(1,2)}$, defined in Eqs.~\eqref{eq:I_rho_1}-\eqref{eq:I_p_2},
characterize the influence of the restricted acceptance angle $\alpha$ on the heating laser.
Clearly, the eigenvalues $\lambda_{\te{n}_i}^{-1}$ of the matrix $\boldsymbol{\Lambda}$ determine the general solution $\vv X(x)$.  The characteristic equation
\(
\left|
  \boldsymbol{\Lambda} - \lambda^{-1} \boldsymbol{1}
\right|
=
0
\)
renders the cubic equation
\(
\lambda^{-3} + a\lambda^{-2} + b\lambda^{-1} + c = 0,
\)
with
\begin{align}
  \label{eq:a_cont}
  a
  &\equiv
  -
  \left(
  \mathcal I_\rho^{(1)}
  +
  \mathcal I_p^{(2)}
  \right)
  \sqrt{2 \mathcal P},
  \\[0.5em]
  \label{eq:b_cont}  
  b
  &\equiv
  2 
  \left(
  \mathcal I_\rho^{(1)}
  \mathcal I_p^{(2)}
  -
  2
  \left[
    \mathcal I_\rho^{(2)}
  \right]^2
  \right) \mathcal P
  -
  1,
  \\[0.5em]
  \label{eq:c_cont}    
  c
  &\equiv
  \mathcal I_\rho^{(1)}
  \sqrt{2 \mathcal P}.
\end{align}
Using the Tschirnhaus-Vieta approach to the solution of cubic equations, one finds that all three roots are real-valued and can be written in the form
\begin{align}
  \label{eq:EV1_cont}
  \lone
  &=
    - \frac{a}{3}
    +
    2 \sqrt{-q}
    \cos 
    \left(
    \frac{\gamma}{3}
    \right),
  \\[0.5em]
  \label{eq:EV2_cont}  
  \ltwo
  &=
    - \frac{a}{3}
    +
    2 \sqrt{-q}
    \cos 
    \left(
    \frac{\gamma}{3}
    +
    \frac{4\pi}{3}
    \right),
  \\[0.5em]
  \label{eq:EV3_cont}  
  \lthree
  &=
    - \frac{a}{3}
    +
    2 \sqrt{-q}
    \cos 
    \left(
    \frac{\gamma}{3}
    +
    \frac{2\pi}{3}
    \right),
\end{align}
where we introduced the auxiliary quantities
\begin{align}
  q
  &\equiv
  \frac{3b-a^2}{9},
  \\[0.5em]
  r
  &\equiv
  \frac{9ab-27c-2a^3}{54},
  \\[0.5em]
  \gamma
  &\equiv
  \arccos
  \left(
    \frac{r}{\sqrt{-q}}
  \right).
\end{align}
The dependence of the eigenvalues $\lambda_{\te{n}_i}^{-1}$ on the particle's propulsion speed (or P\'{e}clet number) and the acceptance angle $\alpha$ are graphically discussed in Sec.~\ref{sec:photon-nudging} of the main text as the analytical expressions above lack an immediate physical insight.

\subsection{2-Species model}
\label{app:2-species-model}

As derived in Sec.~\ref{sec:2-species-model}, the governing equations for the polarization and density profiles within the framework of the 2-state model are structurally equivalent to those of the previously discussed continuous-angle model, namely $\vv X = \boldsymbol{\Lambda}_2 \vv X$. In a dimensionless description (lengths expressed in units of $\sqrt{D/(2k)}$) the matrix $\boldsymbol{\Lambda}_2$ reads
\begin{equation}
  \label{eq:Lambda2_Definition}
  \boldsymbol{\Lambda}_2
  =
  \begin{pmatrix}
    \frac{\sqrt{\mathcal P_2}}{2}
    &
    1
    +
    \frac{\mathcal P_2}{4}
    &
    \frac{\mathcal P_2}{4}    
    \\[0.2em]
    1 & 0 & 0
    \\[0.2em]
    0
    &
    \frac{\sqrt{\mathcal P_2}}{2}
    &
    \frac{\sqrt{\mathcal P_2}}{2}
  \end{pmatrix},
\end{equation}
where we introduced the P\'{e}clet number
\(
\mathcal P_2 \equiv v^2/(2kD)
\)
corresponding to the 2-species model.  The characteristic polynomial
\(
\left|
  \boldsymbol{\Lambda}_2
  -
  \lambda^{-1} \boldsymbol{1}
\right|
=
0
\) delivers the cubic equation
\(
\lambda^{-3} + a\lambda^{-2} + b\lambda^{-1} + c = 0,
\)
with
\begin{align}
  \label{eq:EV1_2spec}
  a
  \equiv
  -\sqrt{\mathcal P_2},
  \qquad
  b
  \equiv
  -1,
  \qquad
  c
  \equiv
  \sqrt{\mathcal P_2/2}.
\end{align}
The solutions are obtained using the same method as in the previous section (Tschirnhaus-Vieta approach).

\subsection{Comparison}
\label{sec:comparison}

\begin{figure*}[tb!]
  \centering
  \includegraphics[width=\linewidth]{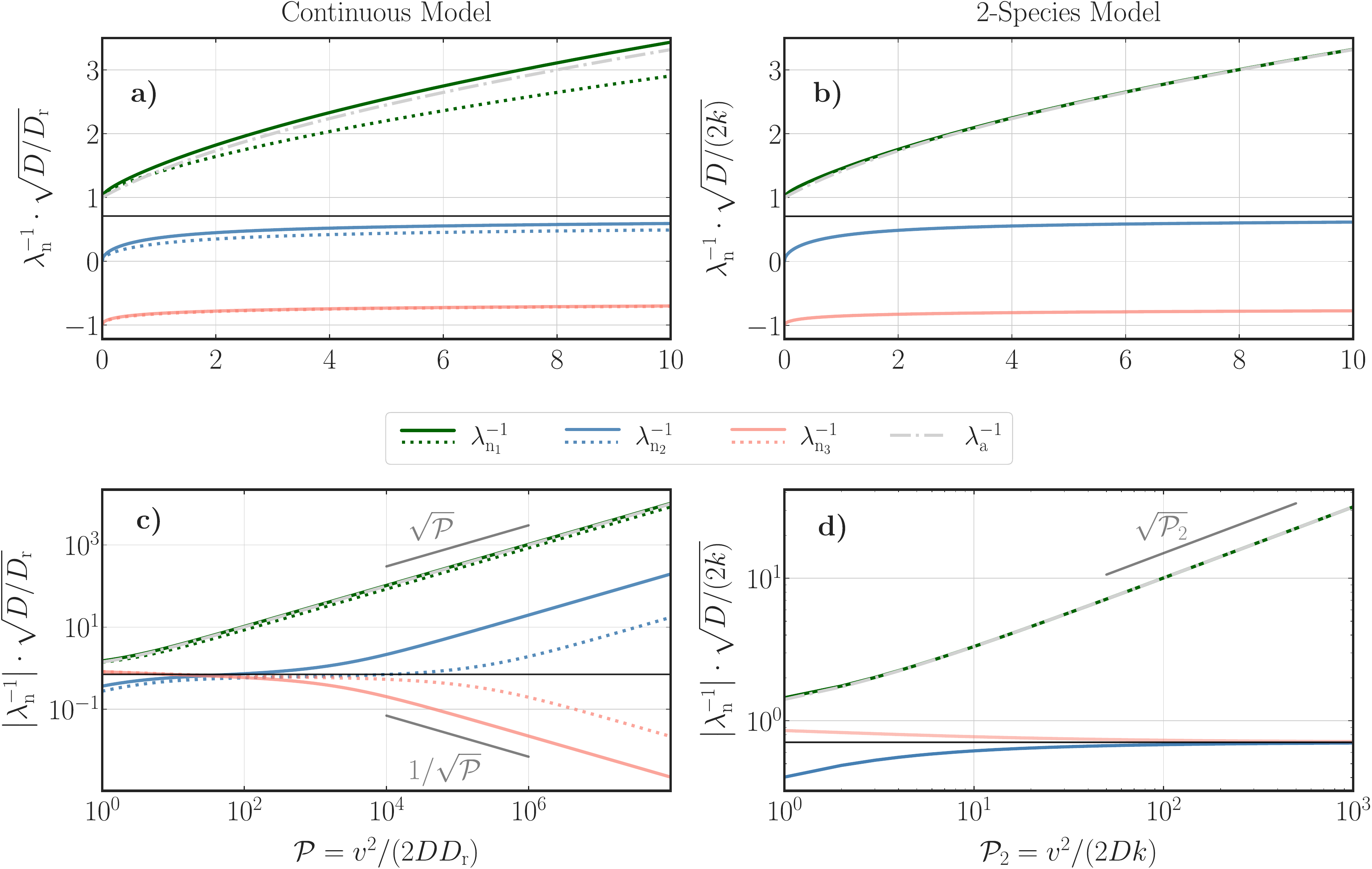}
  \caption{Comparison of eigenvalues. \emph{Left panels}: In \textbf{a)}, Eigenvalues $\lambda_{\te{n}_i}^{-1}$ pertaining to the continuous-angle model are plotted against the P\'{e}clet number $\mathcal P$ using equations \eqref{eq:EV1_cont}-\eqref{eq:EV3_cont} for two fixed acceptance angles, $\alpha = \ang{45}$ (dotted curves) and  alpha = \ang{90} (solid curves). The dashed gray curve corresponds to the inverse length scale (or eigenvalue) $\la^{-1}=\sqrt{1+\mathcal P}$ of a symmetrically active polarization layer. In \textbf{c)}, the absolute value of eigenvalues $\lambda_{\te{n}_i}^{-1}$ are presented on a double-logarithmic scale. \emph{Right panels:} The same eigenvalues pertaining to the 2-species model (cf.~Sec.~\ref{sec:2-species-model}) are plotted against the corresponding P\'{e}clet number $\mathcal P_2$ on a linear-linear [\textbf{b)}] and a double logarithmic scale [\textbf{d)}]. The gray dashed curve corresponds to $\la^{-1}=\sqrt{1+\mathcal P_2}$, the (inverse) natural size of a fully active polarization layer. The  horizontal lines in all four panels correspond to the value $1/\sqrt{2}$.}
  \label{fig:EVs_compare}
\end{figure*}

The upper panels of Fig.~\ref{fig:EVs_compare} compare the eigenvalues $\lambda_{\te{n}_i}^{-1}$ pertaining to the continuous model [a)] to those corresponding to the 2-state model [b)].
We infer that, within the plot range, the eigenvalues of both models display the same qualitative behavior. This observation proves the intuitive conjunction that the 2-species model can serve as a simple model to explain the physics underlying the polarization and accumulation effects.
There are, nevertheless, quantitative and qualitative differences between the two models. This becomes obvious from the lower panels of Fig.~\ref{fig:EVs_compare}, where we plotted the absolute value of the eigenvalues $\lambda_{\te{n}_i}^{-1}$ of both models on a double-logarithmic scale to visualize their behavior for large P\'{e}clet numbers.
We will qualitatively compare both models in the following.  Besides Fig.~\ref{fig:EVs_compare}, we will refer to the content of Tab.~\ref{EVs_limit_behav} showing the limiting behavior of all eigenvalues for low P\'{e}clet numbers.
\setlength{\tabcolsep}{0.5em} 
\begin{table}[tb!]
  \begin{center}
    \begin{tabular}{lll}
      \toprule
      \multicolumn{1}{l}{Eigenvalues}
      &
        \multicolumn{1}{l}{$\mathcal P \ll 1$}
      &
        \multicolumn{1}{l}{$\mathcal P_2 \ll 1$}
      \\
      \midrule
      $\lone$
      &
        $\mathcal I_p^{(2)} \sqrt{\mathcal P/2} + 1$
      &
        $\sqrt{\mathcal P_2}/4+1$
      \\[0.5em]
      $\la^{-1}$
      &
        $\mathcal O(\mathcal P)$
      &
        $\mathcal O(\mathcal P_2)$
      \\[0.5em]
      $\ltwo$
      &
        $\mathcal I_\rho^{(1)} \sqrt{\mathcal P}$
      &
        $\sqrt{\mathcal P_2}/2$
      \\[0.5em]
      $\lthree$
      &
        $\mathcal I_p^{(2)} \sqrt{\mathcal P/2} - 1$
      &
        $\sqrt{\mathcal P_2}/4-1$
      \\
      \bottomrule
    \end{tabular}  
  \end{center}
  \caption{Behavior of the eigenvalues $\lambda_{\te{n}_i}^{-1}$ and $\la^{-1}$ for the continuous model (column $\mathcal P$) as well as the 2-species model (column $\mathcal P_2$) for small P\'{e}clet numbers.}
  \label{EVs_limit_behav}
\end{table}
During the following discussion we will use the notion $\mathcal P_{(2)}$ in order to refer to both P\'{e}clet numbers $\mathcal P$ and $\mathcal P_2$. First, we focus on the eigenvalue $\lone$, which, for low P\'{e}clet numbers, increases proportionally to $\sqrt{\mathcal P_{(2)}}$, irrespective of the underlying model. The (inverse) characteristic size $\la^{-1}$ of a fully active polarization layer grows only as $\mathcal O(\mathcal P_{(2)})$ to leading order in both models. Hence, up to order $\sqrt{\mathcal P_{(2)}}$, the eigenvalue $\lone > \la^{-1}$. On the other end of the spectrum, for $\mathcal P_2 \gg 1$, eigenvalue $\lone \sim \la^{-1} \sim \sqrt{\mathcal P_2}$ for the 2-species model, as can be inferred from Fig.~\ref{fig:EVs_compare} b) and d). The situation is more complicated for the continuous model as can be seen in Fig.~\ref{fig:EVs_compare} a) and c). Depending on the acceptance angle $\alpha$, $\lone$ can be smaller or larger than  $\la^{-1}$, for moderate and large $\mathcal P$.  As numerically determined, choosing $\alpha \equiv \alpha_2 \approx 0.373 \pi$, one has $\lone \sim \sqrt{\mathcal P}$ for $\mathcal P \gg 1$, similar to the the 2-species model. For $\alpha \lessgtr \alpha_2$ one has $\lone \lessgtr \la^{-1}$ for sufficiently large $\mathcal P$.  Note however that the limit $\mathcal P \gg 1$ must be treated with great caution as the orientationally continuous model is based on an approximation \eqref{eq:moment_expansion_general} that looses its justification for large P\'{e}clet numbers. 
Next, we focus on the eigenvalue $\ltwo$. For both the continuous and the 2-species model, $\ltwo$ grows proportionally to the square root of the respective P\'{e}clet number in the case $\mathcal P_{(2)} \ll 1$. As can be inferred from Fig.~\ref{fig:EVs_compare} d), in the limit $\mathcal P_2 \to \infty$, the eigenvalue $\ltwo \to 1/\sqrt{2}$ (dashed line) for the 2-species model. Thus, for infinite activity, the nudging layer decays exponentially (since $\lambda_\te{n_1} \to 0$) over a characteristic length $\lambda_\te{n_2}$ proportional to the extent of a passive layer. This limiting behavior becomes intuitively clear by the observation that one particle species is instantaneously removed from the nudging region ($\lambda_\te{n_1}=0$) while the other species undergoes ordinary diffusion ($\lambda_\te{n_2}\propto \lp$) until its orientation flips.  Regarding the behavior of $\ltwo$ within the continuous model, we refer to Fig.~\ref{fig:EVs_compare} c). The eigenvalue $\ltwo$ first seems to approach a constant value close to $1/\sqrt{2}$ as well, but eventually starts to grow again for further increasing $\mathcal P$. Similar to $\lone$, $\ltwo$ grows proportionally to $\sqrt{\mathcal P}$ for $\mathcal P \gg 1$. Both eigenvalues differ, however, by 2-3 orders of magnitude in this limit, depending on the choice of the acceptance angle $\alpha$.
We emphasize that the limiting behavior of $\lambda_\te{n_2}^{-1}$ for $\mathcal P \gg 1$ is unphysical. At infinite propulsion speed, particles are instantaneously nudged back to the interface as soon as their orientation lies within the acceptance range. The distance covered by Brownian motion until proper re-orientation is proportional to $\sqrt{D/\Dr}$. Therefore, as for the 2-species model, the nudging layer should decay exponentially over said length scale for $\mathcal P \to \infty$.
Figure \ref{fig:EVs_compare} c) shows that the unphysical increase of $\ltwo$ sets in at $\mathcal P \approx 50$-$100$, depending on the choice of the acceptance angle $\alpha$. 
Finally, the eigenvalue $\lthree$ remains negative for all P\'{e}clet numbers $\mathcal P_{(2)}$ within both models. While approaching the value $-1/\sqrt{2}$ in the limit $\mathcal P_2 \to \infty$ for the 2-species model, $\lthree$ approaches zero as $1/\sqrt{\mathcal P}$ for $\mathcal P \gg 1$ within the continuous model. The behavior of $\lthree$ for large P\'{e}clet numbers is unphysical for the same reason as for $\ltwo$.

\end{document}